\documentclass{article}
\usepackage{hyperref}
\usepackage{tikz}
\usepackage{amsmath}
\usepackage{comment}
\usepackage[super]{nth}
\usepackage{xcolor}
\usepackage{svg}
\usepackage{units}
\usepackage[super]{nth}
\usepackage{multirow}
\usepackage{algorithm}
\usepackage{algcompatible}
\usepackage[noend]{algpseudocode}

\algnewcommand{\LeftComment}[1]{\State \(\triangleright\) #1}

\usepackage{enumitem}
\usepackage{amssymb}

\usepackage{listing}
\usepackage{listings}
\usepackage{csquotes}
\usepackage{breakurl}

\definecolor{codegreen}{rgb}{0,0.6,0}
\definecolor{codegray}{rgb}{0.5,0.5,0.5}
\definecolor{codepurple}{rgb}{0.58,0,0.82}
\definecolor{backcolour}{rgb}{0.95,0.95,0.92}

\lstdefinestyle{mystyle}{
    backgroundcolor=\color{backcolour},   
    commentstyle=\color{codegreen},
    keywordstyle=\color{magenta},
    numberstyle=\tiny\color{codegray},
    stringstyle=\color{codepurple},
    basicstyle=\ttfamily\footnotesize,
    breakatwhitespace=false,         
    breaklines=true,                 
    captionpos=b,                    
    keepspaces=true,                 
    numbers=left,                    
    numbersep=5pt,                  
    showspaces=false,                
    showstringspaces=false,
    showtabs=false,                  
    tabsize=2
}

\begin{document}
\sloppy

\date{}

\title{\Large \bf One (Thread) Can Keep a (PRNG) Secret, but not Two }

\author{
{\rm Ehood Porat}\\
Independent researcher
\and
{\rm Amit Klein}\\
The Hebrew University
\and
{\rm Benny Pinkas}\\
Bar Ilan University
}

\maketitle
 
 \begin{center}
 \framebox[8.3cm][s]{
 \begin{minipage}{0.9\textwidth}
 {\large \bf NOTE: This manuscript is based on\\ Ehood Porat's MSc Thesis from 2023.}
 \end{minipage}
 }
 \vspace{1cm}
 \end{center}
 
\begin{abstract}
 We present a novel, practical attack on the IPv6 Fragment ID generation algorithm of XNU, which is the kernel used by Apple products such as macOS and iOS. This attack exploits a \textbf{race-condition} vulnerability in the algorithm's pseudorandom number generator (PRNG) to cryptanalytically break, learn the internal state of the generator, and consequently predict fragment IDs, which, in turn, facilitates an IPv6 fragment spoofing attack. 
 As far as we know, this is the first cryptanalytic attack that is based on exploiting race-conditions. 
 With fragment spoofing, it is possible to partially manipulate UDP datagrams and TCP segments. We showcase a new type of attack on NFS (UDP) where an off-path attacker modifies a file as it is written, and an attack on HTTP (TCP) where an off-path attacker modifies an HTTP request. Apple assigned this vulnerability the CVE identifier CVE-2024-27823 and patched all its XNU-based products against the attack.
\end{abstract}
\section{Introduction}

\subsection{Motivation}
Pseudo-random number generators (PRNGs) are an essential part of network protocols. PRNGs are critical for generating unpredictable identifiers, such as session IDs, sequence numbers, and port numbers. These identifiers serve as unique tokens that enable network protocols to differentiate sessions, maintain the order of messages inside sessions, and protect against connection hijacking, traffic injection, and denial-of-service attacks. The security of network protocols often relies on the unpredictability of the PRNG output. 

Fragmentation is the process of splitting long packets into smaller units that do not exceed the network's maximum transmission unit (MTU). In the context of fragmentation, the importance of unpredictable PRNGs cannot be overemphasized. Predictable PRNGs can introduce significant security weaknesses, enabling attackers to exploit the fragmentation process to hijack traffic or spoof fragments, and potentially disrupt otherwise secure and safe systems. Therefore, ensuring the unpredictability of PRNGs is essential for maintaining the security and reliability of network protocols.

\subsection{Race-Condition Aided Cryptanalysis}
We introduce a new cryptanalysis method that enables an attacker to extract the internal state of a PRNG, which allows the attacker to predict the next PRNG outputs. The new method relies on a race-condition resulting in distinguishable patterns in the PRNG outputs, allowing the attacker to reconstruct the internal state of the PRNG. To the best of our knowledge, this is the \textbf{first attack} that triggers a race-condition in order to preform a cryptanalytic attack, thus opening a door to a new class of cryptanalysis methods.

We target the network PRNG implementation in the XNU kernel, which is a part of Apple's macOS and iOS operating systems. The PRNG algorithm uses a Linear Congruential Generator (LCG) with additional complication layers (including per-output ``injection'' of two random bits from a strong kernel PRNG) to generate a random fragment ID field (32 bits). The PRNG is reseeded every 180 seconds or 1,000,000,000 internal PRNG steps (whichever comes first). As part of the reseeding procedure, the parameters for generating the 31 least significant bits (LSB) are sampled from a strong kernel PRNG.

XNU's implementation of the algorithm (\texttt{ip6\_randomid()}) in the kernel is susceptible to a race-condition that results in duplicate output values. The algorithm is invoked by the IP layer for every outgoing packet that requires fragmentation. Fragmentation can be easily triggered by a non-privileged user process or more importantly, remotely, simply by sending packets whose replies are longer than the MTU. This allows an attacker to trigger the race-condition (i.e. duplicate values) remotely.

\subsection{Overview of the Attack Components}\label{sec:primitiveoverview}
In general, the attack structure that we suggest includes the following components: (1) Triggering a race-condition in the usage of a cryptographic primitive. (2) Using the results of the race condition to break the cryptographic primitive and predict its future outputs. (3) Using these predicted outputs to attack a system.
In the context of our attack against XNU-based devices, these components are as follows:
\begin{enumerate}
\item Triggering a race-condition in the invocation of the fragment ID generation algorithm which generates easily detectable output patterns.
\item A cryptanalytic attack based on the output (detected patterns) of the fragment ID generation algorithm in the previous primitive. The attack results in a full prediction\footnote{up to 4 values} of future fragment IDs until the PRNG's next reseeding.
\item Partial UDP datagram spoofing. Forging an IPv6 second fragment based on the predicted fragment ID, which, when combined with the genuine firstfragment leads to a reassembly of a partially spoofed IPv6 packet.
\end{enumerate}

\subsection{Case Study}
Our case study revolves around the Network File System (NFS) protocol. In particular, we demonstrate that an attacker can inject malicious content into a file on an NFS server which requires Kerberos authentication (RPCSEC\_GSS in the RPC header) and a valid file handle (NFS header) without the attacker possessing valid credentials, by partially spoofing a UDP datagram carrying an NFS WRITE call sent by the NFS client.
It is important to note that our attack cannot be mounted against DNS resolvers/forwarders, which is a typical target of UDP spoofing attacks. The reasons are discussed in detail in \autoref{fragattacktargets}. Additionally, we briefly (due to time and budget constraints) explored the topic of fragmentation in TCP as an additional case study, in which we successfully forced TCP segment fragmentation and esxploited it to modify an HTTP request to include attacker-crafted content.

Enumeration/brute-force attacks on the authentication scheme of NFS (more accurately, on Kerberos) are impractical  as explained in depth in \autoref{rpcsec_gss}. In a nutshell, attacking NFS is equivalent to enumerating at the very least all possible values of a 96-bit cryptographic signature.\footnote{The signature algorithm is HMAC based on truncated SHA1 (aes256-cts-hmac-sha1-96), and it populates the ill-named field ``checksum''.} Those attacks are computationally infeasible and thus do not represent a viable threat.

\begin{figure}[h!]
\centering
\includegraphics[width=222pt,height=254pt]{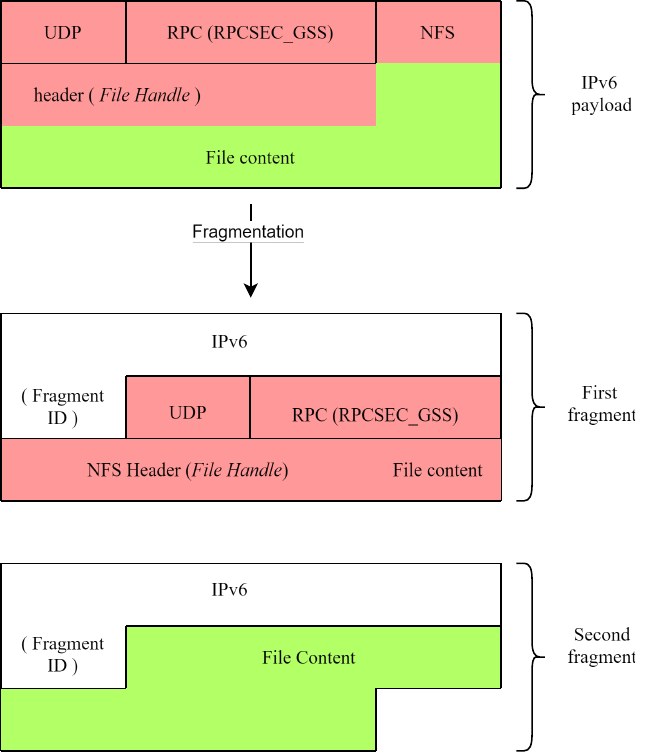}

\caption{IPv6 Fragmentation of an NFS WRITE Call}
    \label{fig:frag}
\end{figure}

In contrast to the above impractical attacks on NFS authentication, we describe a practical attack that we tested extensively. Instead of trying to brute-force the signature, we can exploit the fact that the signature only applies to the RPC header and not the payload of the WRITE call. Considering an NFS WRITE call for a file that is larger than the maximum transmission unit (MTU), which is, e.g., $1500$ bytes for an Ethernet network,\footnote{It is worth noting that while the attack is demonstrated on the WRITE call, it applies to all types of NFS calls/replies that can be fragmented (e.g. READ reply)} and that the default chunk (basic NFS data I/O unit) size for NFS client over UDP in XNU is $8192$ bytes \cite{applenfsmount}, the IPv6 packet will be fragmented. The \nth{1} fragment will contain the UDP, NFS and RPC headers including the signature field in the RPC header, while subsequent fragments will contain only parts of the file content. Thus, an off-path attacker who can predict the fragment ID of the IPv6 packet, can spoof any non-\nth{1} fragment, completely bypassing authentication (see the illustration in Fig.~\ref{fig:frag}).
This is where our primitives (race-condition, cryptanalysis) come into play -- using them, we predict the IPv6 fragment ID value.\footnote{Even when we have the current internal state of the PRNG, the output of the algorithm is not entirely deterministic; rather, at each invocation, the algorithm randomly chooses one next value out of four possible values. But this is a small enough space so the attacker can simply send spoofed fragments for all four values.} With this, we can spoof any non-\nth{1} fragment, bypass Kerberos authentication and spoof the NFS WRITE call.

Another subtle but important requirement is keeping the UDP checksum intact. This can be easily accomplished, assuming the attacker knows the data of the genuine \nth{2} fragment, since UDP's 16-bit checksum is additive. The attacker needs to simply craft the spoofed \nth{2} fragment such that it has the same checksum as the genuine \nth{2} fragment.\footnote{The attacker can choose to target a known file such as a header file of a standard library, predict the genuine \nth{2} fragment and calculate its checksum.}

\subsection{IPv6 Fragmentation Attacks at Large}
The NFS WRITE call spoofing attack we present is a specific example of a generic IPv6 fragment spoofing attack, the latter being so severe an issue that there is, in fact, an RFC called \textit{\enquote{Security Implications of Predictable Fragment Identification Values}}~\cite{rfc7739} dedicated to discussing IPv6 fragmentation attacks. Hence, we consider predicting fragment IDs an important goal by itself.

\subsection{Our Contributions}
\begin{itemize}
    \item The first race-condition-aided cryptanalysis of a real-world system.
    \item A new cryptanalytic attack on the XNU IPv6 fragment ID generation algorithm.
    \item Using the above, a new attack on  NFS's authentication using fragmentation.
    \item Applying the technique proposed in~\cite{pmtudisnotpanacea} to XNU and successfully spoofing TCP segments using our cryptanalytic attack.
\end{itemize}

\section{Related Work}
\subsection{Partial and Complete UDP/TCP Spoofing Attacks}
\subsubsection{UDP Spoofing}\label{sec:rpart}
Spoofing a complete UDP datagram amounts to predicting the 4-tuple connection information: source/destination ports, and source/destination addresses. However, in a realistic attack scenario, spoofing a UDP datagram typically requires knowing only the client side port, as the other values (client and server addresses, server side ``well known'' port) are often trivially predictable. In practice, whether the payload is accepted by the application is a different question. For example, a DNS client requires a correct TXID field (16 bits) in DNS answers. Hence, spoofing a UDP datagram is often limited by a time window and some random or otherwise application identifiers which are not trivially predictable, such as the NFS file handle and the RPCSEC\_GSS signature in RPC. 

An early DNS cache poisoning attack from 2008 that used brute force to enumerate over UDP source ports is~\cite{randomizedports}. UDP port exhaustion attacks were used in~\cite{portex,Alharbi2019CollaborativeCD} though they required running code on the target device and more significantly foreknowledge of the UDP source port. In 2012,~\cite{weakrandom} showed that Android used weak randomness for both the source port and DNS's TXID, but this issue was addressed since then. Another example is~\cite{crosslayerattacksandhowtousethem}, which showed that Linux UDP source port could be predicted, but this issue was addressed since then as well. In contrast, our attack does not attempt to spoof an entire UDP datagram, with the additional merit that, e.g., in NFS, the authentication data is serialized into the \nth{1} fragment (which remains intact) while we spoof only the \nth{2} fragment.

In the literature, there are numerous attacks~\cite{onedomaintorulethemall, domainvalidationformitmresilientpki,poisonovertroubledforwarders,travis,BSDIPIDattack} that partially spoof UDP datagram by taking advantage of the predictability of the IPv4 IPID header field (or its IPv6 equivalent ``fragment ID'' optional header). The IPv4 ID attack presented in~\cite{onedomaintorulethemall} is particularly notable and serves as the basis for many other attacks in this subsection. 

One of the main differences between our attack and~\cite{onedomaintorulethemall} is that we can predict the fragment ID values in advance, whereas if we extend ~\cite{onedomaintorulethemall} which lacks such prediction capability to IPv6, it has to enumerate over the entire IPv6 fragment ID space (up to a factor of 64). This renders the attack impractical for 1Gbps bandwidth.  In addition, the non-random algorithms that ~\cite{onedomaintorulethemall} attacked were fixed long ago by all major operating systems.

In general, IPv4 techniques rely on demonstrating the predictability of fragment IDs and/or on the small ID space, none of which can be directly applied to XNU's IPv6 fragment ID algorithm. The change in ID generation algorithms is also a consequence of multiple papers~\cite{crosslayerattacksandhowtousethem,fromipidtodeviceidandkaslrbypass,offpathtcpexploitsofthemixedipidassignment,travis,BSDIPIDattack} that demonstrated predictable IPIDs in operating systems and how to abuse them for various attacks, prompting operating system vendors to revise and strengthen the IP ID generation algorithms.

\subsubsection{TCP Hijacking}
A partial TCP segment spoofing using IP fragmentation is described in~\cite{pmtudisnotpanacea} and is used to inject a malicious payload to HTTP/BGP. The paper targets fragmentation over IPv4 and assumes that the ID is not randomly generated. As depicted in~\cite[TABLE IV]{pmtudisnotpanacea}, the attack will not work against XNU\footnote{and OpenBSD and FreeBSD} IPv4, because the generation algorithm is not predictable.\footnote{In fact, the attack is also not relevant for IPv6 in Linux  because~\cite{subvertingstatefulfirewallswithprotocolstates} resulted in Linux revising the algorithm for generating fragment IDs to be completely random.} In contrast, we successfully attack XNU's IPv6 fragment ID generation algorithm (a pseudo-random number generator).

A TCP hijacking attack~\cite{offpathtcpexploitsofthemixedipidassignment} (extended in~\cite{offpathtcphijackingattacksviathesidechannelofdowngradedipid})  aims to terminate victim TCP connections or inject forged data into TCP connections by exploiting mixed IPID assignment that was possible in Linux at the time. To achieve this, the attacker ``downgrades'' the IPv4 ID generation algorithm for a TCP connection into a less secure algorithm. Both attacks rely on IPID (i.e. IPv4) hash collision on Linux. However, as mentioned earlier~\cite{subvertingstatefulfirewallswithprotocolstates}, Linux switched to a completely random ID in IPv6, making these attacks ineffective in IPv6. Moreover, in IPv6, a fragment ID extension header is only used when fragmentation is needed. This attack differs from our attack, which targets IPv6 and overcomes XNU's completely random IPv6 fragment ID generation algorithm

In the past some attacks~\cite{Gilad2012OffPathAT,Gilad2014OffPathTI,Gilad2013OffPathHT} used a sequential IPv4 ID as a side channel in order to perform TCP hijacking, those are not relevant nowadays as IDs are generally not sequential. More examples of TCP hijacking via side channel are~\cite{Gilad2013WhenTC,Qian2012OffpathTS,Qian2012CollaborativeTS,Chen2018OffPathTE}. Finally, the attack~\cite{Cao2016OffPathTE} (extended in~\cite{Cao2018OffPathTE}) exploited a side channel in the global rate limit in the challenge ACK mechanism. The attack didn't require code execution on the target but did require binary-search to infer the source port. Mitigation for the attack has been added since then in the form of randomizing the rate-limit count.

\subsubsection{Information Leak}

\paragraph{Data Exfiltration}\label{dataex}
An early example of a covert channel based on predictable IPID from 2008 is~\cite{covertcommunicationsdespitetrafficdataretention}. The attack assumed sequential IPID, which as previously mentioned is not a valid assumption nowadays. In~\cite{subvertingstatefulfirewallswithprotocolstates}, the author exploits information learned from protocol states related to protocol fields e.g. IPID and TCP ISN to establish covert channels. The attack enables devices that are not connected directly to the Internet, to exﬁltrate data out of the network. For exfiltrating data using IPID the author exploited the different generator algorithms' usage of a `Hash-Based Fragment Identification Selection Algorithm'' as recommended in~\cite[section~5.1]{rfc7739}. This enabled the author to coerce a collision and hence learn about the protocol state. While this work is similar to our attack in terms of using protocol state information and performing cryptanalysis on a network protocol header field, it is substantially different in many other aspects. The attack of~\cite{subvertingstatefulfirewallswithprotocolstates} did not predict specific IPID future values -- it merely coerced the state to some distinguishable value. Their cryptanalytic attack was applicable to the IPv6 Flowlabel field and IPv6 fragment ID (of NetBSD) and is fundamentally different from our race-condition based novel approach, due to cryptanalytically crucial difference between the NetBSD and the XNU algorithms. Finally, at the time our research was conducted, the vulnerabilities described in~\cite{subvertingstatefulfirewallswithprotocolstates} were already fixed: Linux switched to a bigger hash table for IP ID generation and a completely random IPv6 fragment ID generation, NetBSD also switched to a better random generation algorithm for Flowlabel and Fragment ID, and Apple released a fix for IPv4 ID, switching to a completely random generation algorithm. 

\paragraph{TCP scan}
The idle-scan technique~\cite{idle} from 1998 exploited sequential IPIDs, which as pointed out earlier is no longer a valid assumption.

\paragraph{Host alias detection and deNATting}
A technique for detecting NATs and counting the number of active hosts behind them was described in 
\cite{atechniqueforcountingnattedhosts}. The technique leveraged two key elements, namely the sequential IPID and the existence of an ID field without fragmentation for a packet. While the second element holds true for IPv4, it does not apply to IPv6 since fragmentation is added as an extension header only when a packet is fragmented. Subsequently,~\cite{ipv6aliasresolutionviainducedfragmentation} proposed a similar attack for host alias detection in IPv6 that also required a sequential fragment ID. Lastly, the attack presented in~\cite{dnsbaseddenatscheme} for client DeNATting necessitated an outgoing DNS over IPv4 query and relied on the same elements as the first attack. All of these are not relevant to XNU as the fragment ID (and IPID) is random and not sequential. 

\paragraph{Traffic interception}
A very early attack~\cite{fragmentationconsideredvulnerable} showed that traffic can be intercepted by overflowing the fragmentation cache if the IDs are predictable. Nowadays such attack would be futile, since IDs are no longer predictable.

\subsection{Cryptanalysis of Fragment-ID/IPID Generation Algorithms}
The work of~\cite{crosslayerattacksandhowtousethem,fromipidtodeviceidandkaslrbypass,BSDIPIDattack,subvertingstatefulfirewallswithprotocolstates} exploited flaws in different PRNGs to predict IPID for different purposes (device tracking, DNS cache poisoning, breaking KASLR). The attacks in~\cite{fromipidtodeviceidandkaslrbypass,crosslayerattacksandhowtousethem} targeted Windows and Linux and were patched in 2019, while~\cite{BSDIPIDattack} targeted IPv4 ID in OpenBSD and was patched in 2007. Our attack targets the current XNU IPv6 fragment ID generation algorithm which has no known vulnerabilities at the time of writing, except the one we disclose. Another fundamental difference (as we already mentioned for~\cite{subvertingstatefulfirewallswithprotocolstates} in \autoref{dataex}) is that previous attacks did not use race-condition to facilitate cryptanalysis.

\subsection{NFS Attacks}
Attacks on NFS generally fall into several categories. \textbf{Downgrade attacks}, described in~\cite{rfc2623,rfc3530,rfc7530}, coerce the NFS client to choose weak security. All such attacks require an attacker with Man-In-The-Middle (MITM) capabilities. \textbf{Spoofing attacks}, described in~\cite{rfc2623,rfc3530,rfc7530,rfc2203} are attacks in which the attacker can modify requests/replies, requiring an attacker that has MITM/eavesdropping capabilities, weak security deployment, or a combination of both. \textbf{Denial of service (DoS)} attacks~\cite{rfc8881,rfc2203} require an attacker with eavesdropping capabilities and/or lack of integrity/encryption/authentication. An impressive security audit of NFS ~\cite{securityauditofnfsv4implementation} lists almost all of the above attacks and renews a theoretical reply attack based on weak NFS cryptography, such as DES and MD5 (that are not commonly used today), and lack of integrity/encryption. In contrast, our attack does not require MITM, eavesdropping, or weak cryptography. Furthermore, our case study is an organizational network that makes these assumptions highly improbable.
It is important to note that~\cite[Figure 5]{securityauditofnfsv4implementation} shows that enabling integrity/encryption has a substantial impact on performance, whereas authentication has minimal impact. This reinforces our assumption that a significant part of the NFSv3 installations has no substantial security measures in place beyond authentication.

\section{The Attack}

\subsection{Attack Overview and Attacker Model}
\label{attackmodel}
\subsubsection{Attack Entities}\label{attackentities}
\begin{figure}[ht!]
  \centering
  \includegraphics[width=220pt,height=128pt]{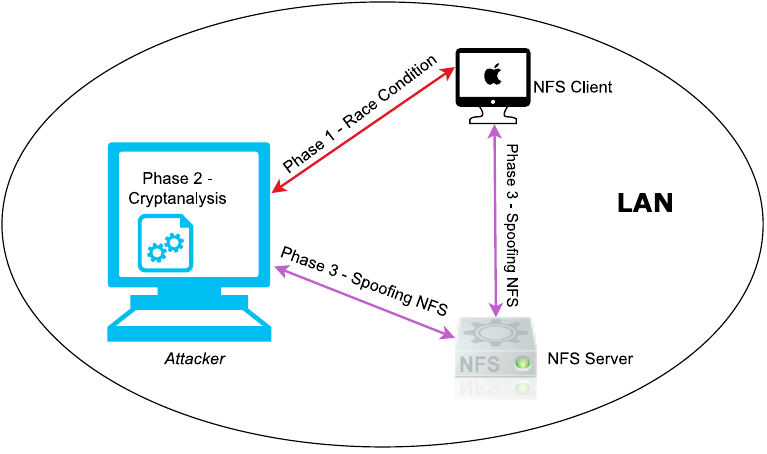}
  \caption{Attack Entities}
  \label{fig:entities}
\end{figure}

The entities involved in the attack are as follows (\autoref{fig:entities}):

\begin{itemize}
\item \textbf{Attacker device}: The entity that carries out the attack. The attacker targets the IPv6 traffic between the macOS client and the server. 
\item \textbf{macOS host}: The attack target (e.g. NFS client, HTTP client) that runs XNU. It should be noted that any XNU-based device may be affected.
\item \textbf{Server}: The server which receives a request from the client, whose non-first fragment is spoofed (e.g. NFS server, HTTP server).
\end{itemize}

\subsubsection{Attack Requirements}
\label{sec:attackrequirements}
\textbf{Attacker device}.
\begin{itemize}
    \item The attacker device must be able to send/receive packets to/from the macOS host and the server. In addition, the attacker device  must be able to send/receive packets to/from the macOS host at a rate of $\approx 1$Gbps in order to trigger the race condition. This condition can be fulfilled by the target device residing on a corporate/ISP network that offers at least 1GBps (or higher) 2-way Internet connection, or having both the target macOS and attacker device reside on the same (modern) LAN.

    \item There is \textbf{no} requirement for the attacker to be able to eavesdrop on the traffic between the macOS client, and the server. 
    \item The attacker device must be able to send packets to the server (e.g. NFS server, web server) with a spoofed source IPv6 address of the macOS host.
    \item The attacker can predict both the payload and the send time of the fragment that he/she is attempting to spoof.
    \item The attacker must be able to associate the responses from the macOS host to individual requests sent by the attacker. This is so that the attacker can process the responses by the sending order of their corresponding requests. This association can be achieved via a simple counter embedded into the requests and echoed back in the responses, any sort of ID embedded in the responses, unique content, or a temporarily unique UDP/TCP request source port. 
\end{itemize}

\textbf{macOS host}.
\begin{itemize}
    \item The macOS client must run a  ``fragmentation service'' (see below) or a ``malicious app'' (a low privileged attacker controlled process). 
\end{itemize}

Given the above requirements, the attacker is able to predict the next fragment ID, and hence spoof any non-\nth{1} fragment of UDP (and TCP). The attack has two variants, in both the attacker leverages his/her ability to predict fragment IDs, to forge fragments originating from the client and intended for the server. The race condition is triggered by having the client send fragmented packets concurrently. The attacker can trigger the race condition via two methods: by a ``fragmentation service'' running on the client, or by a local app running on the macOS client (or iPhone). These two variants are described below. 
Next, the attacker runs a real-time cryptanalysis process with the duplicated fragment IDs as input. The output of this process is the current internal state of the PRNG, which can be used to predict future fragment IDs.

The two attack variants are as follows:

\textbf{The ``fragmentation service'' variant} 
In this attack model, the macOS host runs a ``fragmentation service'' that listens and sends packets back concurrently. This can be any benign service that happens to exhibit the required behavior. The attacker device sends requests to the fragmentation service on the macOS host, and the macOS host responds with fragmented IPv6 packets. The attacker then observes the fragment IDs to detect the race-condition. In our experiments, due to time shortage, we used a simple echo server over UDP to model the fragmentation service. 

\textbf{The ``malicious app'' variant}.
A \textit{malicious app} is an attacker-controlled low-privilege process running on the macOS host (e.g. malware,  or a malicious app if the device is an iPhone; for the ease of testing we used macOS as our showcase)), which sends packets using the standard socket API to the attacker device. The malicious app simply sends, as rapidly as it possibly can, packets (UDP datagrams or TCP segments) that get fragmented, to the attacker device. Here too, the attacker device is only required to observe the fragment IDs.

Finally, while our main use case is NFS over UDP (\autoref{nfsusecase}), we also managed to elicit fragmentation in TCP based on the work in \cite{pmtudisnotpanacea} and successfully modified an HTTP POST request (over TCP/IPv6) sent from a macOS host to a web server.

\subsubsection{Use Case: NFS}\label{nfsusecase}
The use case we demonstrate targets a macOS host running an NFS client, that uses an NFS server for file backup, all within an organization network. The attack involves three entities, namely, the attacker device (a device on the organization network that is fully controlled by the attacker), the macOS host (NFS client), and the NFS server. The attacker device spoofs a fragment that is sent from the macOS host to the NFS server. The NFS server thus receives the spoofed information.

The NFS server is configured with a security flavor that requires Kerberos authentication provided by RPCSEC\_GSS (described below). This configuration was intentionally selected to demonstrate an attack scenario against a secured NFS setup. Kerberos is vastly common in organizations, hence the obvious choice of a security level for the NFS server. It is assumed that the organization uses NFSv3 over UDP 
as it is still widely used\cite{loadbalancerNSFv3}, and there is no clear performance advantage to moving to NFSv4\cite{martin2008,nfsv3isfaster1,nfsv3isfaster2}. Additionally, from a security perspective, the main advantage of NFSv4 is that Kerberos support is mandatory \cite{nfssecurityinbothtrustedanduntrustedenvironments}, but since the corporate already employs Kerberos on the NFS server, there is no urge to upgrade from NFSv3 to NFSv4.

In order to elicit fragmented traffic from the macOS host, we deploy a straightforward UDP echo server as a``fragmentation service'' (described below) on the NFS client. In reality, any benign service that can be forced to send fragmented IPv6 datagrams can serve as a ``fragmentation service''. Alternatively, the attacker can run a local low-privileged malicious app on the NFS client to achieve the same goal.

The attacker device must be able to spoof the source IPv6 address of the macOS host when communicating with the NFS server. The attacker device must also be able to capture the fragmented traffic sent to it from the NFS client. The attacker device must be able to send/receive at a rate of $\approx 970$Mbps to the NFS client. See illustration in \autoref{fig:entities}.

It is assumed that the macOS is about to write a file to the NFS server, by issuing an NFS WRITE call with the file contents. The attacker can target any file whose content is partially known (specifically the attacker must know the full content of a non-first fragment, an example is given in  \autoref{app:headerspoof}), preferably one that would compromise privacy or integrity if tampered with. The aim of the attack is to maliciously modify the content of a file that is being written to the NFS server. The attack assumes that the NFS WRITE call is fragmented, which occurs for every file size that exceeds the MTU. The attacker has the freedom to modify the content within any non-first fragment. However, there are two restrictions: the size of the modified content cannot be changed, and two bytes must be reserved for correcting the checksum (more than two if there are constraints on the characters used). 
In the use case we describe, the attacker targets a header file named \texttt{random.h}, which belongs to an open-source library that is publicly available. We emphasize that this choice serves as an illustrative example and in practice it could be any file fully known to the attacker. For the sake of simplicity, we assume that this file is the sole item being copied and altered during the attack.

\paragraph{``fragmentation service''}\label{sec:echoserver}
In order to observe the desired special sequences that indicate a race-condition (\autoref{trigger}), the attacker first needs the target macOS host to emit fragmented traffic (since the IPv6 fragment ID is only present in fragmented packets). To this end, a simple C++ ``echo server'' is used, as described in this section. This server is run on the target macOS host, and models a UDP service that is accessible to other nodes (such as the attacker). Unfortunately, due to time constraints, we could not find a native service that operates in a similar manner.
The echo server code consists of one listener thread that en-queues each incoming request into a single shared, thread-safe input queue, and $N$ worker threads, where $N$ is the number of cores available on the macOS device. Each worker thread de-queues an available request and sends it back to the source address. All threads are bound to the same port but use different sockets. 

\paragraph{NFS configuration }\label{nfs_config}
The NFS protocol supports multiple modes of security, namely ``security flavors''. The ``security flavors'' are configured on the NFS server. While the default flavor is  ``sys’’,  it is not considered secure \cite[section-2.2.1]{rfc2623}.  Since the most secure flavor is  RPCSEC\_GSS \autoref{rpcsec_gss},  we have deliberately chosen to demonstrate an attack scenario against NFS using the RPCSEC\_GSS flavor, and specifically when using Kerberos authentication.
The NFS server in our case, is a macOS server (distinct from the macOS target) running the built-in macOS NFS server software as described in \autoref{attackmodel}, though any NFS server may be used.

\subsubsection{Attack Procedure (the NFS Use Case)}

The attack goal is to inject malicious content into the backup copy of the file \texttt{random.h} while en-route from the macOS target (NFS client) to the NFS server. This is achieved by partially spoofing a UDP datagram, the one used for the NFS WRITE call that transmits the file \texttt{random.h} from the NFS client to the NFS server. The attack is based on three main primitives as described in \autoref{sec:primitiveoverview}.

The information leakage as will be described in \autoref{triggers} exploits a race condition in the PRNG, detectable by duplicate fragment IDs, effectively  allowing an attacker to gain insight into the PRNG's internal state. By leveraging this leaked information, the attacker can preform cryptanalysis and predict subsequent fragment IDs, as will be detailed in \autoref{cryptanalysis}. This prediction enables the spoofing of non-first fragments with malicious content, thereby bypassing Kerberos authentication, as will be discussed in \autoref{nfscraft}.

\subsubsection{Use Case: HTTP} 
The use case we demonstrate involves a macOS host operating an HTTP client. In this scenario, the attacker spoof a fragment to alter the HTTP request body. The simplicity of this use case is intentional, as it aims to showcase the attack over TCP. This differs from the NFS use case, where the focus is on circumventing a security feature.

The attack procedure is similar to the NFS use case, with the necessary alterations.

\subsection{XNU IPv6 ID Generation Algorithm}
\label{randomipid}
XNU's IPv6 fragment ID field (32 bits) is populated using the X2/32 PRNG (following the naming convention introduced in \cite{BSDIPIDattack}). The ID field comprises a most significant bit (MSB) part, and 31 least significant bits part. The field is computed using a PRNG that we describe herein. The PRNG is reseeded using the \textproc{Reseed} function of \autoref{alg:XNU-ID} every 180 seconds or 1,000,000,000 internal PRNG steps (whichever comes first). As part of the reseeding procedure, the parameters for generating the least significant 31 bits are sampled from a strong kernel PRNG, and the MSB is flipped. Since the MSB is static between reseeding events, it can be ignored for the remainder of the discussion, and we will henceforth use the term ``the PRNG'' or ``the algorithm'' to denote the algorithm that generates the least significant 31 bits of the IPv6 fragment ID, defined by the \textproc{Generate-ID} function of \autoref{alg:XNU-ID}. The algorithm is designed to produce unique outputs during two consecutive reseeding periods, which fulfills the basic requirement for IPv6 IDs (uniqueness over short periods of time). However, as we note in App.~\ref{app:collisions}, this property does not strictly hold due to (very rare) possible collisions.

Note that in a single PRNG invocation used to generate the IPv6 fragment ID, the PRNG draws a random number $1 \leq n \leq 4$ (using a different, cryptographically-strong PRNG) and runs $n$ internal steps.
\begin{algorithm}[ht]
\begin{algorithmic}[1]
\caption{XNU IPv6 ID Algorithm (31 Least Significant Bits)}
\label{alg:XNU-ID}
\LeftComment{ $M=1836660096=2^7 \cdot 3^{15}$}
\LeftComment { $N=2147483629$ (prime), $\varphi(N)=N-1=2^2\cdot 3^2 \cdot 59652323$}
\Procedure{Reseed}{}
\State $x \gets \Call{random}{\{0,\ldots,M-1\}}$ 
\State $s_1 \gets \Call{random}{\{0,\ldots,2^{31}-1\}}$  
\State $s_2 \gets \Call{random}{\{0,\ldots,N-2\}}$  
\State $a \gets 49^{\Call{random}{\{0,\ldots,2^{31}-1\}}} \mod M$  
\State $b \gets \Call{random}{\{x|0<x<2^{32} \land \gcd(x,M)=1\}}$
\State $g \gets 2^{\Call{random}{\{x|0<x<N \land \gcd(x,N-1)=1\}}} \mod N$
\EndProcedure
\Procedure{Generate-ID}{}
\State $n \gets \Call{random}{\{1,\ldots,4\}}$
\For{$i=1$ to $n$}
\State $x \gets (ax+b) \mod M$
\EndFor
\State return $s_1 \oplus (g^{x \oplus s_2} \mod N)$ 
\EndProcedure
\end{algorithmic}
\end{algorithm}

\subsection{Primitive 1 -- Triggering a Duplication} \label{triggers}

\subsubsection{Triggering a Race Condition}\label{trigger}
Reviewing the source code of XNU we found that no lock serializes the access to the IP layer, hence no thread-safety is guaranteed by the kernel for the PRNG.

We used two alternative methods for triggering a duplication:
\begin{enumerate}
    \item Remotely, by using a ``fragmentation service'' described in \autoref{sec:attackrequirements} (specifically the echo server described in \autoref{sec:echoserver}).
    \item Locally, by running an un-privileged simple program written in C (\textit{malicious app}) that launches $c$ threads where $c$ is the number of cores, with each thread sending out UDP datagrams rapidly. Note that the only requirement is the ability to send large UDP datagrams very fast.
\end{enumerate}

The traffic in both methods is captured on the attacker device (remotely). In our experiments, we use Tshark with a custom Lua tap for performance. 

\subsubsection{Race Condition Analysis}\label{sec:specialseq}
Over the next few paragraphs, we are going to describe the following:
\begin{itemize}
\item An analysis of the race-condition.
\item The ``special'' sequence ($\mathit{XYZY}$) and its relation to the race-condition.
\item A hypothesis on the connection between the ``special'' sequence and the internal state of the PRNG, including an empirical measurement to assess our hypothesis.
\item The results of the empirical measurements.
\end{itemize}
Throughout this subsection, \textit{step} numbers refer to lines in  \autoref{alg:XNU-ID}.

XNU has a single, global PRNG instance for generating fragment IDs used every time the kernel fragments IPv6 packets. The x64 and ARM64 implementations of XNU have the CPU load the current state of the PRNG ($x$) from memory just before the main PRNG loop (i.e., before step 12), and store the new state to memory at the end of the loop (after step 13).
The loop itself is carried out purely in the CPU using registers. One can verify that in both architectures (x64 -- \autoref{xnu_ipv6_disasm} lines 2-5, ARM64 -- \autoref{xnu_ipv6_disasm_arm64e} lines 2-9) the values (keys and state) from the global PRNG object \texttt{randomtab32} are loaded into the registers at the start of the algorithm. Afterwards, no memory is accessed until line 51 (x64) and line 39 (ARM64) where the new state is written back from a register (\texttt{edx} (x64), \texttt{W9} (ARM64)) into \texttt{ru\_x} which is a part of \texttt{randomtab32}.

The XNU implementation of the PRNG is not thread-safe, i.e. there is no explicit locking mechanism that serializes the access to the PRNG internal state and ensures its integrity. It is possible to force two kernel threads to invoke the PRNG algorithm concurrently, and thus it is possible to end up in the following scenario where the current internal state $x$ is right after the last output $X$: 
\begin{enumerate}[label=\roman*]
    \item Thread $A$ enters \textproc{Generate-ID}, draws $n_A>1$ and reads the internal state $x$ (before step 12) into its registers (part of its thread context) 
    \item Thread $B$ enters \textproc{Generate-ID}, draws $n_B<n_A$ and reads the internal state $x$ (before step 12) into its registers
    \item Thread $A$ calculates the next internal state ($y$) from its registers by running the loop $n_A$ times, stores it in the PRNG global state (RAM) after step 13, and emits the PRNG output $Y=s_1 \oplus (g^{y+s_2} \mod N)$ (step 14). 
    \item Thread $B$ calculates the next internal state ($z$) from its registers by running the loop $n_B$ times, stores it in the PRNG global state (RAM) after step 13, and emits the PRNG output $Z=s_1 \oplus (g^{z+s_2} \mod N)$ (step 14). 
    \item An arbitrary consumer (Thread $A$, Thread $B$ or any other thread) consumes an additional PRNG value, by reading the internal state $z$, drawing $n=n_A-n_B$ and calculating the next state which is exactly $y$ (because $n_A=n_B+n$) and outputting $Y=s_1 \oplus (g^{y+s_2} \mod N)$.
\end{enumerate}

The net result is the following PRNG output sequence (we extend the sequence to include the previous output $X$ of the PRNG, which corresponds to $x$):
$$\mathit{X,Y,Z,Y}$$
This sequence can be easily identified by an observer -- it contains two identical (i.e. \textbf{duplicate}) values ($Y$), separated by a different value ($Z$), with yet another different value ($X$) preceding them.

First, for clarity of discussion, let us define, for an internal state $x$, its offset from the seed (i.e. the number of PRNG inner loop steps) as $s_x$. The cryptanalysis requires (a lot of) pairs of outputs $(X,X')$ whose respective pairs of internal states $(x,x')$ satisfy $(s_x \mod 4) \ne (s_{x'} \mod 4)$. The cryptanalysis can withstand a certain amount of error, i.e. we do not need 100\% of the pairs to be correct. 

Therefore, when the above described race-condition occurs it results in: 
\begin{enumerate}
    \item A stream of outputs $\mathit{XYZY}$ where $X,Y,Z$ correspond to the internal states $x,y,z$
    \item Since the internal state $x$ is followed by $z$ and also by $y$ (where $s_y>s_z$) and in this case  $s_x < s_z < s_y \leq s_x+4$, we obtain the two inequalities $s_y \mod 4 \neq s_z \mod 4$, and $s_x \mod 4 \neq s_z \mod 4$.
\end{enumerate}

We denote 
$p_{(s_x,s_z)}$ as $\mathit{Prob}(s_x \ne s_z \mod 4|\mathit{XYZY})$, $p_{(s_z,s_y)}$ as $\mathit{Prob}(s_z \ne s_y \mod 4|\mathit{XYZY})$. 
Per our hypothesis, had it only been possible to get $XYZY$ through the particular race condition event $s_x < s_z < s_y \leq s_x+4$, then we would have had $p_{(s_x,s_z)} = p_{(s_z,s_y)}=1$. However in reality, there are multiple fundamentally different race condition events that lead to $XYZY$ and therefore $p_{(s_x,s_z)}$ and $p_{(s_z,s_y)}$ are slightly lower than 1 (but still considerably higher than 0.75). 

Since empirically $p_{(s_z,s_y)}$ is slightly larger than $p_{(s_x,s_z)}$ we denote by $p$ the average of $p_{(s_z,s_y)},p_{(s_x,s_z)}$. We hypothesize that $p$ would be significantly higher than 0.75 (whereas when observing an arbitrary sequence $XYZ$,  $Prob(s_x \ne s_z \mod 4|\mathit{XYZ}) \approx 0.75$ since it is the same as $\mathit{Prob}(\text{random two bits} \neq \text{random two bits})$) and likewise $Prob(s_z \ne s_y \mod 4|\mathit{XYZ}) \approx 0.75$). 

In order to prove this hypothesis, one would have to instrument the kernel. Aside from the practical difficulties of instrumenting XNU, the instrumentation would damage the natural course of the race-condition, defeating the original purpose of the instrumentation. Therefore, we empirically prove this intuition (this proof is not part of the attack, it is merely a set of lab experiments).

\label{definitive1} We look at cases of $\mathit{XYZY}$, and calculate the portion of ``good'' pairs (for the cryptanalysis). Formally,  for a sequence $\mathit{XYZY}$ and internal states $\mathit{x,y,z}$ corresponding to $\mathit{X,Y,Z}$, the experiment that is described below calculates $p$ (the average of $\mathit{Prob}(s_x \mod 4 \neq s_z \mod 4|\mathit{XYZY})$ and $\mathit{Prob}(s_z \mod 4 \neq s_y \mod 4|\mathit{XYZY})$). We then show that $p$ is significantly higher than 0.75 which justifies the use of this approach for our cryptanalysis.

In order to compute $s_x$ for each fragment ID, we first dump the kernel memory before each experiment to get the initial PRNG state. Then, independently from the kernel, we advance the PRNG algorithm from its initial in single steps (by setting $n=1$ in step 11), and for each fragment ID $X$, we search for the matching algorithm output $X$ (whose internal state is $x$) and record the number of PRNG steps from the initial state (as $s_x$). 

We conducted multiple experiments, in each we counted the number of occurrences of the sequence $\mathit{XYZY}$ -- denote this by $P$, and using the $s_x$ values, we counted the number of inequalities among the two pairs in each occurrence of the  sequence $\mathit{XYZY}$ -- denote this by $I$. Each instance of the sequence can contribute up to two inequalities as described earlier, so we have $E(I)=p \cdot 2P$. Hence we estimate $p=\frac{I}{2P}$.

Results on x64 architecture from a 2012 MacBook device, yielded $p$ values ranging at~$ 0.9024 < p < 0.9261$. Results from  ARM64 architecture from a 2021 MacBook device, yielded $p$ values ranging at~$ 0.8138 < p < 0.8388$.

The experiment proves our hypothesis that $p$ is significantly higher than 0.75, which is an enabler for the attack.

\subsection{Primitive 2 -- Cryptanalysis}\label{cryptanalysis}
Before diving into our cryptanalysis primitive, it should be noted while there are other attacks against this PRNG, as described in Appendix~\ref{app:impracticalattacks}. However, those attacks are impractical, either due to their excessive run time and/or due to their negligible coverage of the key space.

\subsubsection{Overview}
We extract the static key of the PRNG ($s_1,g,s_2,a,b \mod M$) in four phases. In the first phase, we extract $s_1$ using our fundamental insight regarding the higher-than-average probabilities of some inequalities on internal state bits when the special fragment ID sequence $XYZY$ is observed. In the second phase, we use $s_1$ from the first phase to extract $g$. This time we use pairs of consecutive fragment IDs, exploiting the dependencies in their internal state bits. In the third phase we use the extracted $s_1, g$ to extract $s_2$ by extending the technique of the second phase. With $s_1,g,s_2$ we can calculate the internal state $x$ of the PRNG corresponding to every fragment ID we observe. Finally in the fourth phase, given the states of consecutive fragment IDs 
we extract $a,b \mod M$ by solving two linear equations (mod $M$) over two unknowns $a, b$.

\subsubsection{Notations and Partial Internal States}
The PRNG defined in~\autoref{alg:XNU-ID} has a static key $a, b, g, s_1, s_2$ and a secret dynamic internal state $x$ that is modified with each PRNG invocation, and is initialized with another part of the key $x_0$. We remind that ~\autoref{alg:XNU-ID} also defines $M$ and $N$ to be specific constants.

We write $x^{(m)}$ as a shorthand for $x \mod m$. Particularly, for a {\em full} internal state $x$, we define a {\em partial internal state (modulo $m$)} $x^{(m)} \equiv x \mod m$. It is then easy to see the following:
\begin{itemize}
    \item If $m|M$, and $y=ax+b \mod M$ then $y^{(m)} = a^{(m)}x^{(m)}+b^{(m)} \mod m$. 
    \item If $m|48$ then $a^{(m)}=1$ (this is by construction). Therefore, applying the linear transformation ($x^{(m)} \mapsto a^{(m)}\cdot x^{(m)} + b^{(m)}$) $n$ times would result in $x^{(m)} \mapsto x^{(m)} + n \cdot b^{(m)} \mod m$ for any $m|48$.
    \item If $m=2^i3^j$ then $b^{(m)}$ is invertible modulo $m$ (since by construction $b \mod 2 \neq 0, b \mod 3 \neq 0$).
\end{itemize}
We also define a map $\log_2:\{1,\ldots,N-1\} \mapsto \{0,\ldots,N-2\}$ as follows: 
$\log_2(2^{n} \mod N) \mapsto (n \mod (N-1))$.
This map is well defined because 2 is a generator of $\nicefrac{\mathbb{Z}}{N\mathbb{Z}}$.
\subsubsection{The Basic Insight}
\label{sec:basicinsight}
We now look at the partial internal states of the special sequences ($\mathit{XYZY}$) we considered  in~\autoref{sec:specialseq} 
with $m=4$, i.e., the two least significant bits of the internal state. Recall that $s_z-s_x$ denotes the number of times the PRNG applies the transformation from the partial internal state $x$ to $z$, i.e. $z^{(4)} = x^{(4)}+(s_z-s_x) \cdot b^{(4)} \mod 4$. Since $Prob(s_z-s_x \neq 0 \mod 4|\mathit{XYZY})=p$ and $b$ is invertible $\mod 4$ we have $Prob((s_z-s_x) \cdot b^{(4)} \mod 4 \neq 0|\mathit{XYZY})=p$, i.e. $Prob(x^{(4)} \neq z^{(4)}|\mathit{XYZY})=p$. Therefore (henceforth, all inequalities hold with probability $p$ given $\mathit{XYZY}$):
$$(x \oplus s_2)\mod 4 \neq (z \oplus s_2)\mod 4$$
We can multiply each side by $\log_2 g$ (which is invertible modulo $N-1$ by $g$'s construction, and since $4|N-1$, it is also invertible modulo 4),\footnote{Note that $8 \nmid N-1$, which is why we cannot extend this approach beyond the two least significant bits.} so the following holds:
$$(\log_2 g)\cdot(x \oplus s_2)\mod 4 \neq (\log_2 g)\cdot(z \oplus s_2)\mod 4$$
We can rewrite the above into:
\begin{gather*}
(\log_2 (g^{x \oplus s_2} \mod N))\mod 4 \\
\neq \\
(\log_2 (g^{z \oplus s_2} \mod N)) \mod 4
\end{gather*}
And finally, since $X=s_1 \oplus (g^{x \oplus s_2} \mod N)$ and
$Z=s_1 \oplus (g^{z \oplus s_2} \mod N)$
we get:
$$(\log_2 (X \oplus s_1))\mod 4 \neq (\log_2 (Z \oplus s_1)) \mod 4$$
And likewise with $Y$ and $Z$ since $Prob(s_y \neq s_z \mod 4|\mathit{XYZY})=p$.

In other words, if we observe a PRNG sequence $\mathit{XYZY}$ then we have a probability $p$, significantly higher than 0.75, for each of the following equations to hold, \textbf{for the correct $s_1$}:
\begin{equation}
\label{eq:ineq}
\begin{aligned}
(\log_2 (X \oplus s_1))\mod 4 \neq (\log_2 (Z \oplus s_1)) \mod 4 \\
(\log_2 (Y \oplus s_1))\mod 4 \neq (\log_2 (Z \oplus s_1)) \mod 4
\end{aligned}
\end{equation}
Whereas for a random $s_1$, the probability is $\approx 0.75$. This is the basic insight that lets us conduct an efficient cryptanalysis of the XNU algorithm.
The cryptanalysis is composed of four phases. The first phase is detailed here, while the remaining phases are described in \autoref{app:cryptanalysis} for the interested reader.

\subsubsection{Cryptanalytic Attack Phase 1 -- Extracting \texorpdfstring{$s_1$}{s1} }\label{phase1}
Offline, the attacker prepares a $\log_2(\cdot) \mod 4$ table for $\{1,\ldots,N-1\}$, i.e., a table where $(2^{n} \mod N) \mapsto (n \mod 4)$.

In the online phase, the attacker collects $XYZY$ sequences (denote their number by $P$). Then the attacker goes over all $2^{31}$ possible values for $s_1$. For each value, the attacker counts how many~\autoref{eq:ineq} inequalities hold among the $\mathit{XYZY}$ sequences -- designate this by $Q$. The attacker picks the $s_1$ value that maximizes $Q$ as the presumably correct $s_1$.

For the correct $s_1$, $E(Q)=p\cdot2P$ (as we explain in~\autoref{sec:specialseq}). For a random $s_1$,  $Q\sim\mathit{N}(\frac{3}{4}\cdot 2P,\frac{3}{16}\cdot2P)$.

To reliably eliminate $2^{31}-1$ false positive $s_1$ values,  $k=7$ standard deviations suffice, thus we require
$Q\geq \frac{3}{2}P+k\sqrt{\frac{3}{8}P}$.

This yields a lower bound on $P$, since we want (say) at least half of the phase 1 attempts to succeed:
$$ E(Q)=p\cdot 2P \geq \frac{3}{2}P+k\sqrt{\frac{3}{8}P}$$
Therefore, even for the ideal case $p=1$, $P\geq 937$. 

This phase takes $2^{31}\cdot 2P$ calculations and ends up with the correct $s_1$. With this $s_1$, the attack proceeds to the next cryptanalytic phases (as described in \autoref{app:cryptanalysis}).

\subsubsection{PRNG Synchronization Step}
\label{syncphase}
In order to be predict the next fragment IDs, the attacker has to be synchronized with the macOS host (i.e. the internal state must be advanced the same number of times). De-synchronization can happen due to packet loss, e.g. the attack experienced packet loss and did not receive the last fragment sent by the macOS host. In order to synchronize to the PRNG current state, the  attacker device sends one more request to the ``framgnetation service'' on the macOS host. The attacker extracts the fragment ID from the reply, and computes its internal state $f$ using $s_1, g, s_2$ obtained via the cryptanalysis. Finally the attacker copies the internal state $f$ into his/her PRNG simulator which can produce the next fragment IDs.

\subsection{Primitive 3 -- Partial UDP Datagram Spoofing}\label{nfscraft}
\subsubsection{NFS protocol}
In order to understand our attack, a short explanation about the NFS WRITE call is required. NFS is built on top of ONC-RPC (~\cite{rfc5531}). NFS calls cannot be serialized directly to a UDP datagram, and instead are serialized as ONC-RPC calls.

\paragraph{RPC call}\label{rpcsec_gss}
The NFS protocol supports multiple modes of security. The most secure is  RPCSEC\_GSS. Underlying RPCSEC\_GSS is the Generic Security Standard API (GSS-API) for authentication, integrity, and privacy typically implemented via Kerberos. The RPCSEC\_GSS is provided as fields inside the RPC call header, therefore its fields precede the NFS WRITE call fields.

\begin{enumerate}
    \item Xid: A 32-bit unique transaction ID for the RPC call, used to match the response to the corresponding request.
    \item Credentials: Authentication information for the RPC call. In the context of RPCSEC\_GSS the relevant fields are:
    \begin{enumerate} 
        \item A 4-byte handle received from the RPC server.
        \item A 4-byte sequence number that must be within a specified sequence window exchanged during context creation.
    \end{enumerate}
    \item Verifier: This field is used for data consistency checking. In our case, the verifier contains an a cryptographic keyed-hash digest (12 bytes) of the call header from the xid up to and including the credentials.~\footnote{The signature algorithm in our case is HMAC-SHA1-96}
\end{enumerate}
An attacker who wants to spoof a WRITE call to an NFS server operating in \texttt{sec=krb5} mode, as in our case, has two options. The first is to attempt to falsify a Kerberos ticket~\footnote{which, in our case, uses the AES256-CTS encryption algorithm}. This approach is not feasible as Kerberos's underlying cryptography is generally considered secure.

The second option is to attempt to enumerate the 4-byte handle that identifies the Kerberos ticket. However, this approach is not practical either, as the attacker would need to ensure that the sequence number falls within the allowed window. In XNU, the allowed window is hard-coded to 256 (\cite{applenfsgss} \texttt{GSS\_SVC\_SEQWINDOW}), which means that the search space is effectively $2^{32} \times (\frac{2^{32}}{256})$. Additionally, the attacker would need to compute the correct HMAC over the RPC header up to and including the credentials field ($2^{96}$ possible values), making this method completely unfeasible.

\paragraph{Crafting the NFS WRITE call.}
In order to successfully execute the spoofing attack as outlined in the introduction, the attacker needs to know, in advance (1) the genuine file's content that is being written to the target server; and (2) the size of the (genuine) \nth{2} fragment.
The attacker can target any file that is fully known in advance. Consider the following example: a developer wants to backup the code he works on, so he sets up a periodic task to copy it to a remote server. The code contains some publicly available header files (.h), e.g. header files for public \nth{3}-party libraries. Once the backup task runs, the attacker injects malicious code to one of the header files. Eventually the project will be compiled with the embedded malicious code, and subsequently the code could be run locally or deployed to numerous other devices. 

To demonstrate the attack, consider the header file located at \texttt{\$\{SDK\_HOME\}/usr/include/sys/random.h} which is provided by Apple's ``CommandLineTools''. This file is public, and as such, it fulfills the \nth{1} requirement. The \nth{2} requirement can be fulfilled by taking into account the file size and the total size of the headers.We demonstrated the attack with a realistic payload as described in~\autoref{app:headerspoof}.

\paragraph{UDP checksum.}\label{sec:checksum}
To successfully spoof the \nth{2} fragment of the NFS WRITE call, it is imperative to ensure that the UDP checksum is correct. Since the algorithm for computing the checksum in UDP is additive, the correction process is relatively straightforward. All that is required is to append a spare word (two bytes) at the end of the spoofed \nth{2} fragment and utilize it to adjust the checksum to that of the genuine \nth{2} fragment, before sending the spoofed fragment.

\section{Experiments and Results} 
\subsection{Experiment}\label{fullexperiment}

\paragraph{Cryptanalytic backend.}\label{machinelimitation}
The attacker's ``cryptanalytic backend'' is an abstraction for a powerful compute device controlled by the attacker, which is used for the cryptanalysis work load. Conceptually the cryptanalytic backend can be thought of as part of the attacker entity. It has no limitation on the physical location, i.e. if the attacker device is powerful enough (e.g. a device with many cores or perhaps multiple GPUs), it can serve as the cryptanalytic backend. Alternatively, if the attacker controls multiple devices on the local network, then they can collectively serve as the cryptanalytic backend, and the attacker can distribute the work load among the individual devices. Finally, the cryptanalytic backend can be completely remote -- on or off a public cloud. It should thus be stressed that the physical separation we describe in the proceeding is merely due to a weak device on our setup, and that in other circumstances, the cryptanalytic backend can be completely local.
In order to conduct the cryptanalysis, we utilized a virtual machine on Azure (Standard\_D96as\_v4) with 48 cores and 384GiB RAM as our cryptanalytic backend. 
Using this backend machine, our cryptanalytic attack takes 4-15 minutes. This does not suffice to mount a real-time attack, because the PRNG is reseeded every three minutes. As described in the~\autoref{attackmodel}, the attack has two variants, using local app running on the macOS host, or using a ``fragmentation service'' that runs on the macOS host. The race condition phase takes $\approx 30\text{s}$ on the former and $\approx 70\text{s}$ on the latter. This leaves even less time for real-time cryptanalysis.

With access to a backend machine that is six times (for the app variant), or eight times (for the  ``fragmentation service'' variant) faster, it would have been possible to run the cryptanalysis process in real-time. 
To validate the effectiveness of our attack, in spite of not being able to run it in real time, we developed two methods:
\begin{enumerate}
    \item \textbf{Re-seed time extension}, in which we artificially extend the re-seed time using a kernel extension we wrote. We modify the reseed  parameter (\texttt{ru\_out}) in the \texttt{randomtab32} PRNG state to read 3600 seconds (arbitrarily chosen to be high enough), instead of 180 seconds. This allows us to complete the cryptanalysis, and use the PRNG state thus revealed to mount the attack.
    \item \textbf{Retrospective state comparison}, in which we read the current internal PRNG state after the race condition phase from the kernel memory, mount the attack using it (instead of using the cryptanalytic result) and retrospectively compare the PRNG state extracted with the results obtained through cryptanalysis carried out after the fact.
\end{enumerate}

\paragraph{Step 1: Experiment setup.}
The experiment involves four participants.
\begin{itemize}
    \item \textbf{Attacker (two parts).} The attacker in the experiment comprises two logically separate parts, which on our setup are also physically separated. 
    The \textit{frontend} is an attacker device connected to the switched-LAN. Henceforth  we will use the term ``attacker'' or ``attacker device'' to denote the frontend.
    The \textit{backend} is the attacker's cryptanalysis backend machine on Azure.
    
 \item \textbf{macOS host (NFS client).} A Macbook connected to the same switched-LAN as the attacker device. The macOS host runs an echo server as a model of the ``fragmentation service''. 
    \item  \textbf{NFS server.} A second Macbook  connected to the same switched-LAN as the attacker device. (As explained in~\autoref{nfs_config}, the operating system of the server can be arbitrary. We happened to use MacOS.)
    
\end{itemize}

Before the experiment starts, the macOS host (NFS client) mounts an NFS share locally at \texttt{/Users/user/nfs\_share} and remotely on the NFS server at \texttt{/Users/center}. The ``echo server'' is started on the macOS host, listening on UDP port $3333$.

\paragraph{Step 2: Re-seed timeout extension.}
 As explained above, we extend the re-seed timeout to 3600 seconds using a kernel extension to allow real-time cryptanalysis. The re-seed timeout  parameter takes effect only after a re-seeding event, i.e. once we modify the \texttt{ru\_out} parameter, we need to wait for the next re-seeding event before we start the experiment. All subsequent re-seeding events will have a timeout of 3600 seconds. The modification was required in order to mount the attack in real time due to our compute time constraints. In addition we ran a ``retrospective'' experiment as detailed in \autoref{sec:NFS_results}, which retained the original re-seed timeout.

\paragraph{Step 3: Identifying the reseed time.}
Since the MSB is static between reseeding events (as described in~\autoref{randomipid}), the attacker can identify the next reseed event (within 180 seconds, which we prolonged to 3600 seconds ) by observing a flip in the MSB of fragment IDs. In order to maximize the attack time window, the attacker sends one request per second to the ``fragmentation service'' (in our model this is the ``echo server'') of the macOS host. 
Each echo request has a payload size of 1500 bytes (the Ethernet MTU is 1500, thus the overall IPv6 length of the request is higher than the MTU), therefore, the request and the reply are  fragmented. The attacker captures each reply and extracts the MSB of the fragment ID, and compares it to the MSB of the first reply. After at most 180 seconds, the attacker will identify a bit-flip indicating that a reseed occurred.

Once the attacker detects a reseed event, he/she can start the attack with a guaranteed almost 180 seconds time window in which the PRNG keys remain intact. 
(Due to our limited compute power we extended this time window to 3600 sec.)

\paragraph{Step 4: Race condition phase.}
As described in~\autoref{attackmodel} the attacker is uses a simple four bytes index at the start of the UDP payload. The UDP payload has 1500 bytes (the Ethernet MTU is 1500, thus the overall IPv6 length of the request is higher than the MTU). 

\begin{enumerate}
    \item  The attacker runs a listener (even a simple one such as netcat) on the UDP source port (on the attacker device) that he/she uses to send packets to macOS host, in order for the attacker device OS not to send ICMP port unreachable messages back to the macOS host.
    \item The attacker device  then starts a ``tshark'' (sniffer) that captures the fragment IDs and the sender ordinal from the replies. 
    \item The attacker device sends the first $L+1$ packets (i.e.  $2(L+1)$ fragments) that were pre-prepared to the ``fragmentation service'' (i.e. to UDP port 3333) on the macOS host, with a short delay of 5 milliseconds before each packet, to assure they arrive and processed in order. The fragment IDs in the replies from the macOS host are then used in~\autoref{app:phase2},  therefore the more accurate the order of those packets is -- the less noise we will have. In practice we use $L=600$ (601 packets in $\approx 3$ seconds).
    \item The attacker sends the following  $5\times 10^6 - (L+1)$ packets that were pre-prepared to the  ``fragmentation service'' (i.e. port 3333 over UDP) on the macOS host as fast as possible. In our experiment, this takes $\approx 67$ seconds.
\end{enumerate}

\paragraph{Step 5: Cryptanalysis.}
The attacker gets from the ``tshark'' (sniffer) the fragment IDs and their sender ordinal. Then, the attacker extracts only the necessary data for cryptanalysis, by following this procedure:
\begin{enumerate}
    \item Reordering the fragment IDs using the sender ordinal.
    \item Generating an extract of the data, composed of the following: \begin{enumerate}
        \item The first $L+1$ packets.
        \item All the ``special'' sequences ($\mathit{XYZY}$). We denote the number of such sequences in the data $P$.
        \item The last fragment ID is to be used for synchronizing the attacker's PRNG state with the current  state of the target's PRNG.
    \end{enumerate}
\end{enumerate}
The data thus extracted is sent to the cryptanalytic backend. The technical details of communicating with the cryptanalytic backend and the data size and bandwidth estimation are provided in ~\autoref{app:inputsize}.

The cryptanalytic backend  receives the processed fragment IDs from the previous step, performs the cryptanalysis and returns the output (the current internal state of the PRNG, i.e. the \texttt{randomtab} struct) to the attacker device. 
The attacker sends one additional echo request to the ``fragmentation server'' on the macOS host as described in the ``PRNG synchronization phase''~\autoref{syncphase}. Once the attacker's PRNG state is synchronized, the attacker computes the next four fragment IDs.

\paragraph{Step 6: Spoofing.}
In order to create valid fragments that can be accepted by the NFS server, the attacker uses the known content for the genuine $\nth{2}$ fragment NFS WRITE call to create a spoofed $\nth{2}$ fragment with a matching UDP checksum (as described in~\autoref{sec:checksum}). We used two alternative payloads: inert content (\enquote{\texttt{THIS\_IS\_WORKING\_\it{random\_hex\_chars}}}) and weaponized content which demonstrates our ability to trojanize C code (\autoref{app:headerspoof}). 

The attacker then sends four copies of the spoofed \nth{2} fragment, each with one of the four fragment IDs he/she computed in the previous step. Assuming no other organic fragments are received by the NFS server (which may compete over the fragment cache with the attacker's fragment), there is a 60 seconds window starting when the spoofed \nth{2} fragment is sent by the attacker 
for the genuine \nth{1} fragment to arrive and be assembled with the spoofed \nth{2} fragment.  If, during this window, the macOS host's NFS client writes to the NFS share, e.g. executing the command \\
\texttt{cat /Users/user/secret\_project/random.h > /Users/user/nfs\_share/ \\ random.h}, the attack succeeds, as the server will immediately match the newly arrived genuine \nth{1} fragment with the spoofed \nth{2} fragment (still in the cache), and the genuine \nth{2} fragment that arrives shortly afterwards ``loses the race'' and is not used.

\paragraph{Step 7: Attack success verification.}
We retrieve the file \texttt{/Users/center/random.h} from the NFS server and check that our content is injected successfully.

\subsection{Cryptanalysis Results}\label{cryptresults}
We computed the following for each experiment in the context of the cryptanalytic attack.
\begin{itemize}
\item $P$ -- Number of occurrences of the sequence $\mathit{XYZY}$.
\item $I$ -- The maximum number (over all $s_1$ candidates) of inequalities among the two pairs in each occurrence of the  sequence $\mathit{XYZY}$. Each instance of the sequence can contribute up to two inequalities, as described in ~\autoref{sec:basicinsight}.
\item $E, \sigma$  -- Per ~\autoref{phase1}, for random $s_1$, the number of inequalities is distributed $\sim\mathit{N}(\frac{3}{4}\cdot 2P,\frac{3}{16}\cdot2P) = \mathit{N}(E,\sigma^{2})$.
\item $z= \frac{I - E}{\sigma}$ -- The distance between $I$ and $E$ in units of $\sigma$.
\end{itemize}

$z$ provides an estimation of how likely the cryptanalysis is to extract $s_{1}$. Recall that in phase 1 (\autoref{phase1}) we choose $s_1$ that maximizes the number of inequalities, i.e. that produces $I$ inequalities, and  we require it to be at least $k=7$ standard deviations above the average $E$ for random candidates. 

In all the experiments in the ``fragmentation service'' variant (implemented as an echo server), the cryptanalysis was successful, with $z$ values in the range $8.7 < z < 12.2 $,  i.e. $z \gg 7$ which satisfies the requirement. In the ``app'' variant all experiments concluded in a successful cryptanalysis as well with even  better $z$ values in the range $ 12.7 < z < 15.5$.

\paragraph{Cryptanalysis verification against the ground truth.}
The validation of the cryptanalysis is performed by comparing the extracted state of the PRNG to the cryptanalysis output ($s_1, s_2, g, a, b$). The  $s_1, s_2, g, a$ fields should match exactly, whereas for the value of $b$, $b \mod m$ should match. We ran 10 experiments, in all of them the cryptanalysis results matched perfectly the PRNG state we extracted from the RAM.

\subsection{NFS Attack Results}
\label{sec:NFS_results}
\begin{figure*}[ht!]
  \centering
  \includegraphics[width=4.8in]{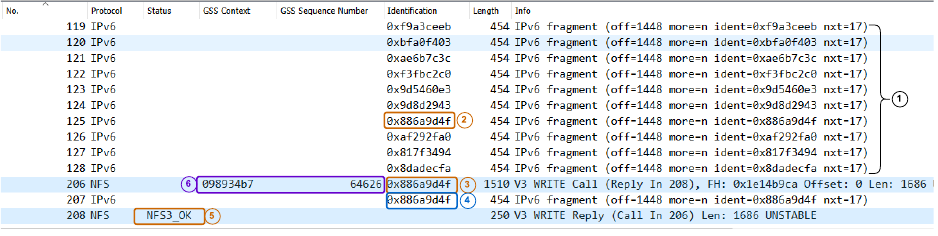}
  \caption{Attack flow. The spoofed $\nth{2}$ fragments (we send more than four fragments in order to sync with the current state of the PRNG; in later experiments we used a better approach as described in~\autoref{syncphase} ) \textcircled{1}, among them the spoofed $\nth{2}$ fragment \textcircled{2} with the correct fragment ID. The macOS host (NFS client) request is fragmented into \textcircled{3} (which contains the authentication data \textcircled{6}) and \textcircled{4}. The NFS server responds with an NFS success status \textcircled{5}.}
  \label{fig:attackcapture}
\end{figure*}

\begin{figure}[H] 
  \includegraphics[width=4.8in]{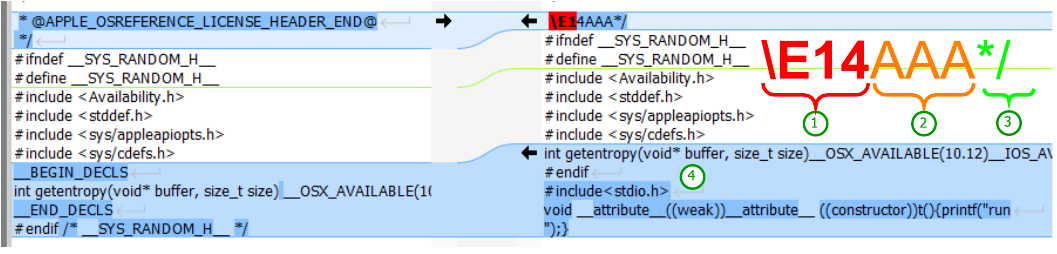}
  \caption{Side-by-side file comparison for the target of the attack. The left-hand side is the original file, and on the right-hand side is the modified file. \textcircled{1} the spare word (E1 04) we used for correcting the checksum. \textcircled{2} padding we used to match content length. \textcircled{3} The enclosing comment characters, the spare checksum word is non-ASCII, in order for the .h file to compile we must be inside a comment. \textcircled{4} Injected trojan code that runs whenever the compiled program is executed.}
  \label{fig:diff}
\end{figure}

We ran the re-seed time extension variant of the experiment  10 times with random strings\\
(\enquote{\texttt{THIS\_IS\_WORKING\_\it{random\_hex\_chars}}}) and succeeded to inject the string to the written file every time.
The column ``server hash'' displays the hash of the file after the modification (\autoref{tab:nfsattackresults}). We also conducted the experiment using a realistic payload, an example of such modification is shown in~\autoref{fig:diff}.
The flow of the attack can be seen in the packet capture in~\autoref{fig:attackcapture}.

We ran similar experiments using the ``retrospective state comparison results'' method in order to verify that changing the reseed time does not affect the essence of the attack. As expected, all 10 experiments succeeded, which clearly demonstrates that extending the PRNG reseed timeout does not alter the system in a way that affects the attack (other than extending the attack window). 

\begin{table}[ht]
\resizebox{\columnwidth}{!}{
\begin{tabular}{|l|l|l|l|l|}
\hline
index & success & P & cryptanalysis time & server hash \\ \hline
1 & True & 1653.0 & 07m 00s & dd4f3a57b859eca8b31a15f7cfcc6622 \\ \hline
2 & True & 1735.0 & 04m 34s & c793d7380f789c09a11212fa9dc5d92a \\ \hline
3 & True & 1807.0 & 04m 49s & 76d240407ab4bbffdc10090921e1783b \\ \hline
4 & True & 1734.0 & 03m 53s & 72e0f1c8d07564bc83fb9ab6b42ad5e2 \\ \hline
5 & True & 1771.0 & 08m 09s & 3d7563a7992050feaa982b7a94c8a075 \\ \hline
6 & True & 1717.0 & 02m 32s & 1f1c8265c5caa8cbd8f53a60f0c8ea53 \\ \hline
7 & True & 1677.0 & 07m 17s & 3999ba99bce6752519ace4d0bfd8dec7 \\ \hline
8 & True & 1703.0 & 06m 48s & 3ada22fdd30bef554279c92b21e778f3 \\ \hline
9 & True & 1740.0 & 06m 09s & 5a2fa14d6923db7cc2bb9a5f6e352c59 \\ \hline
10 & True & 1734.0 & 04m 51s & cb8b68dbd48453fbfac5cc90f16bd3ee \\ \hline
\end{tabular}
}
\caption{NFS attack for MacBook Air (2021), model MGN63HB/A (Apple M1 CPU, 8 cores, 8GB RAM,  macOS 13.3 (22E252), kernel version xnu-8796.101.5~3.}
\label{tab:nfsattackresults}
\end{table}

\subsection{Tested macOS/XNU Versions}
The combination of the cryptanalysis phase and the race condition phase was tested on the hardware and software combinations listed in~\autoref{tab:versions}. This demonstrates that our attack works across different architectures (x64, arm64) and many kernel versions.

\begin{table}[ht]
\resizebox{\columnwidth}{!}{
\begin{tabular}{|l|l|l|l|l|}
\hline
CPU & architecture & XNU version & macOS version & Major \\ \hline
Apple M2 chip & arm64 & 8792.81.3$\sim$2 & 13.2.1 & Ventura \\ \hline
Apple M1 chip & arm64 & 10063.101.15$\sim$2 & 14.4 & Sonoma \\ \hline
Apple M1 chip & arm64 & 8796.101.5$\sim$3 & 13.3 & Ventura \\ \hline
Apple M1 chip & arm64 & 8792.81.2$\sim$2 & 13.2 & Ventura \\ \hline
Apple M1 chip & arm64 & 8792.41.9$\sim$2 & 13.0 & Ventura \\ \hline
Apple M1 chip & arm64 & 8019.80.24$\sim$20 & 12.2.1 & Monterey \\ \hline
Apple M1 chip & arm64 & 7195.121.3$\sim$9 & 11.4 & Big Sur \\ \hline
Intel i7-3820QM & x64 & 6153.141.28.1$\sim$1 & 10.15.7 & Catalina \\ \hline
\end{tabular}
}
\caption{Tested HW/SW versions}
\label{tab:versions}
\end{table}

\subsection{HTTP Experiments}
The HTTP experiments only differ from the NFS experiments in a few steps which are described in \autoref{app:http}. The attack is carried out by spoofing the body of HTTP POST request.

\section{Discussion}

\subsection{Fragmentation and DNS}\label{fragattacktargets}

The DNS protocol used to be a the typical target for UDP fragmentation attacks. However, the ``DNS Flag Day 2020'' \cite{dnsflagday2020} addressed the operational and security problems in DNS caused by fragmentation. The DNS flag day recommended limiting accepting DNS messages over 1232 bytes. This means that the overall IPv6 length is limited to 1280, which is the lower limit for MTU in IPv6. Therefore, DNS servers do not accept fragmented packets over IPv6. Thus, any attack based on fragmentation for DNS cache poisoning cannot be mounted against modern DNS servers (since circa 2021-22).

 \subsection{Attack Effectiveness}
As shown in \autoref{tab:versions} our attack is very flexible and works in various setups, on completely  different architectures (i.e. x64 and ARM64), and against different XNU versions (8796.101.5$\sim$3, 8792.81.2$\sim$2, 8792.41.9$\sim$2, 8019.80.24$\sim$20, 7195.121.3$\sim$9, 6153.141.28.1$\sim$1) and macOS versions (Sonoma, Ventura, Monterey, Big Sur, Catalina). Overall the attack proved to be very stable as we managed to mount it without modification for all of the above operating systems and hardware.

Our attack has two variants for triggering the race condition, namely the ``malicious app'' and the ``fragmentation  service''.  We demonstrated that both variants can successfully be used for cryptanalysis in \autoref{cryptresults}. 
We observed 
$P$ in the range $1653-1807$ for the ``fragmentation  service'' variant and $2757-2909$ for the ``malicious app'' variant, both clearly fulfill $P \geq 937$ (the lower bound calculated in \autoref{phase1}).

\subsection{Attack Impact}
The attack we demonstrated can facilitate several nefarious activities. One immediate result is code execution, which can be achieved by modifying any file that contains code, e.g. .h/.c/.cpp for C/C++ projects, .py for python projects, or any shell script, etc. We showcased an injection of a generic trojanized code into a C/C++ header file (.h) which will run whenever the library/binary compiled from it is loaded into memory. The technical details of how to trojanize C/C++ header files are described in \autoref{app:headerspoof}.

Our attack can also be used for modifying configuration files, which can be very lucrative targets. For example, modifying a  server configuration file can compromise the entire system. And since it is very common to leave most of the configuration file in its default form, the attacker can leverage this to predict the second fragment size and content when the file is transferred and alter it in transit. For example, an HTTPS server configuration modification could potentially have serious security consequences such as changing the certificate used for TLS/SSL, changing the privileges required for routes, changing the web-root to a controlled path, and even code execution by loading external modules. Another example might be modifying any documents the client stores for backup, e.g. HTML files. An attacker can modify an HTML file so that it collects payment card and account credentials (phishing-like), or to serve malware to users when the HTML file is served as part of a website.

An alternative scenario is where we target the NFS READ call and achieve similar effects, this time for files served from the NFS server to its clients. It is also trivial for an attacker to inflict a denial-of-service condition, as all he/she has to do is to send four fragments with the four possible next fragments IDs, with an invalid checksum or with an incorrect length field and the genuine packet will get discarded.

Lastly, it is possible to attack application traffic over TCP. An obvious  target would be to modify an HTTP POST request to preform authenticated operations. Another prime target is FTP, where we achieve over TCP the same goals as with NFS. 

\subsection{Root Cause}
\label{rootcause}
``normal'' thread-safety violation bugs are notoriously difficult to spot due to their dependence on very specific conditions that are hard to predict a-priori and to reproduce/simulate. However, since these bugs do have an observable impact, with sufficient time and testing, such bugs do manifest themselves and are eventually fixed. We show here that this is not necessarily the case with PRNG thread-safety violation (race condition) bugs. These are not flagged because users and testers are not aware of the subtle security implications of such bugs, which otherwise do not affect the performance of their systems. So ``no one complains'', and the vulnerabilities remain in the system possibly for over a decade. To wit, the vulnerability we exploit here is 12 years old.\footnote{To the best of our understanding, this vulnerability was introduced in xnu-1699.32.7, which was released on October 4, 2012.}

\section{Conclusion}

Our work exposes the severe effects that race-conditions have on PRNGs. It demonstrates that a race-condition can facilitate a practical cryptanalytic attack that would have been otherwise impossible.  
Our attack on the widely popular XNU kernel leverages a race condition to perform a cryptanalytic attack that results in the complete extraction of the internal state of the PRNG used by the fragment ID generation algorithm. This enables us to spoof IPv6 fragments carrying either TCP segments (based on \cite{pmtudisnotpanacea}), or UDP datagrams.  Our comprehensive experiments demonstrated the stability and reliability of such attacks even on different hardware/architectures and software versions. While our attack's target is the PRNG of XNU which is not ``industrial strong'' per-se, the attack concept itself might be applied on industrial strength cryptographic algorithms as well.

\section*{Ethics considerations}
\subsection*{Vendor status}
We disclosed the vulnerability to Apple on February \nth{28}, 2024. On July \nth{29}, 2024, Apple completed fixing the vulnerability in all its operating system releases and assigned it the vulnerability identifier CVE-2024-27823.\footnote{https://nvd.nist.gov/vuln/detail/CVE-2024-27823}

\clearpage

\bibliographystyle{plainurl}

\clearpage
\appendix

\section{Cryptanalysis Overview}\label{app:cryptanalysis}
\subsection{Phase 2 -- Extracting \texorpdfstring{$g$}{g} (and three least significant bits of \texorpdfstring{$s_2$}{s2})}\label{app:phase2}

From this phase onward, the attacker does not need the $\mathit{XYZY}$ sequences. In this phase, the attacker uses simple pairs of consecutive PRNG outputs. 

Offline, attacker prepares a $\log_2(\cdot)$ table for $\{1,\ldots,N-1\}$, i.e., a table wherein $(2^{n} \mod N) \mapsto (n \mod (N-1))$. 

    In the online phase, the attacker calculates $\log_2(F \oplus s_1)$ for all PRNG output values ($F$) the attacker has. Then, the attacker enumerates over all possible $g$ values ($\varphi(N-1)=715827864 \approx 2^{29.4}$ values) and over the three least significant bits of $s_2$. We choose to focus on the three least significant bits because this is the minimal number of bits that can be used for elimination (using four bits would have worked as well, but it requires enumerating over one more bit).

Note: due to some arithmetic properties, this phase always results in eight indistinguishable candidate keys. For the three least significant bits of $s_2$, we get four combinations, and for $g$ there are two candidates: the correct value, and its inverse, $g^{-1}$. These false positives are passed to the next phase, but this property can also be used in this phase to enumerate three less bits: two less bits of $s_2$, and one less bit of $g$. For simplicity, we ignore this nuance: we describe the full enumeration of $g$ and the three least significant $s_2$ bits and assume a single candidate is passed to the next phase. 

\textbf{Phase 2 Elimination:} The attacker uses $L$ pairs of consecutive outputs (we explain how $L$ is chosen later). 
Given a guess of $g$ and $s_2 \mod 8$, and for each PRNG output $X$ (which is part of a pair), the attacker can obtain a partial internal state $x^{(8)}$: $x^{(8)} = ((\log_2(X \oplus s_1)) \cdot (\log_2 g)^{-1} \mod 8) \oplus (s_2 \mod 8)$.

For a pair of consecutive partial internal states ($x^{(8)},x'^{(8)}$), $x'^{(8)} = x^{(8)} + n \cdot b^{(8)} \mod 8$, with $1 \leq n \leq 4$. Thus $x'^{(8)}-x{(8)} \mod 8 \in \{b \mod 8,2b \mod 8,3b \mod 8,4b \mod 8\}$. Since $b$ is odd, $b \mod 8 \in \{1,3,5,7\}$. Thus, the attacker sets up 8 counters, $C_0,\ldots\,C_7$, and increments $C_{x'^{(8)}-x^{(8)} \mod 8}$ for each ($x^{(8)},x'^{(8)}$) pair.

Define $S_1=C_1+C_2+C_3+C_4, S_3=C_3+C_6+C_1+C_4, S_5=C_5+C_2+C_7+C_4, S_7=C_7+C_6+C_5+C_4$. For the correct key, all pairs (up to some noise) will contribute to $S_{b \mod 8}$, thus one of the sums will be close to $L$. For a random $g$ and $s_2$, $C_i\sim\mathit{N}(\frac{1}{8}L,\frac{7}{64}L)$, thus $S_j\sim\mathit{N}(\frac{1}{2}L,\frac{7}{16}L)$. To reliably eliminate $2^{29.4+3}=2^{32.4}$ false keys, we require $k=7$ standard deviations above the expected value in one of the sums, that is, for some $i$:
$$S_i \geq \frac{1}{2}L+k\sqrt{\frac{7}{16}L}$$
This also induces a lower bound on $L$, since $S_i \leq L$: $L \geq 86$. More pairs are needed if there is noise (false pairs).

This phase takes $2^{32.4}L$ calculations, but with the optimization described in the above note, this can be reduced to $2^{29.4}L$.

\subsection{Phase 3 -- Extracting the Remaining \texorpdfstring{$s_2$}{s\_2} Bits}
The attacker enumerates over the remaining 28 bits of $s_2$, and for each candidate, the attacker can now calculate the full internal state $x$ corresponding to the output $X$: $x  = (((\log_2(X \oplus s_1)) \cdot (\log_2 g)^{-1}) \oplus s_2$. 
Since $x<M$, a first quick elimination is to verify that for every output on the key candidate in question. This check is not affected by noise, as long as the output does indeed come from the same PRNG seed.

The main phase processing proceeds similarly to the previous phase. The attacker considers $L'$ pairs of consecutive outputs, and for each pair, calculates the internal states ($x,x'$). Now the attacker applies modulo $m=48$ (the maximal $m$) on the internal states.\footnote{$m$ needs to be chosen such that $m|48$, and also $m>4$. In our code, we chose the less optimal $m=12$ for ease of programming.}  

Following the approach of the previous section, the attacker sets 48 counters, $C'_0,\ldots,C'_{47}$, and increments $C'_{x'-x \mod 48}$ per each pair. Recall that $b$ is invertible modulo 2 and modulo 3 by construction, hence $b \mod 48 \in \{1,5,7,11,\ldots,47\}$. The attacker therefore defines $S'_1=C'_1+C'_2+C'_3+C'_4,S'_5=C'_5+C'_{10}+C'_{15}+C'_{20},S'_7=C'_7+C'_{14}+C'_{21}+C'_{28},\ldots,S'_{47}=C'_{47}+C'_{46}+C'_{45}+C'_{44}$. Then for the correct key, $S'_{b \mod 48}$ will have a value almost $L'$ (up to noise). For a random key, on the other hand, $C'_i\sim\mathit{N}(\frac{1}{48}L',\frac{47}{48^2}L')$, so $S'_j\sim\mathit{N}(\frac{1}{12}L',\frac{47}{576}L')$. The attacker needs to eliminate $2^{28}$ false keys (actually, per the note in the previous section, this figure is actually $2^{31}$), so again, we require one of the sums to be at least seven standard deviations above the expected value for a random key, i.e., $S'_j \geq \frac{1}{12}L'+7\sqrt{\frac{47}{576}L'}$.
As usual, this induces a lower bound on $L'$ since $S_j \leq L'$. However, the usual lower bound calculation is not applicable here because $L'$ is small. Instead, we estimate the lower bound as follows: in order to have a false positive in a particular sum, its counters must be hit every time, and this happens at probability $\frac{1}{12^{L'}}$. Since this may happen at any of the 16 sum variables $S_j$, the overall probability for a false positive at a particular key is $\frac{16}{12^{L'}}$, so we require:
$2^{31}\frac{16}{12^{L'}} \ll 1$, so, say, $L' \geq 12$ satisfies the requirement. 

This phase requires $2^{28}L'$ calculations for a single incoming ($s_1, g, s_2 \mod 8$). In practice, there are 8 incoming candidates, so the total calculations required is $2^{31}L'$.

\subsection{Phase 4 -- Extracting \texorpdfstring{$a$}{a} and \texorpdfstring{$b$}{b}}
At the beginning of this phase, the attacker has the correct key fields $s_1,g,s_2$, and thus the attacker can calculate the full internal state for every output value. Moreover, from phase 3, the attacker also has $b \mod 48$, which is the index of the sum that exceeds the threshold. With this, the attacker can calculate $n$ for every pair ($x,x'$): since $x' \mod 48 = x + b \cdot n \mod 48$, we have $n=(x'-x)\cdot b^{-1} \mod 48$. The attacker picks several pairs with $n=1$. Every pair contributes a linear equation for $a,b$ over $\nicefrac{\mathbb{Z}}{M\mathbb{Z}}$: $x' = ax+b \mod M$. Since two equations suffice to extract $a,b \mod M$, the attacker can solve sets of randomly chosen two equations, and rank the result $(a,b \mod M)$. The key ($a,b \mod M)$ that is ranked highest should be the correct key.

This phase requires a negligible number of calculations. At the end of this phase, the attacker has $s_1,g,s_2,a,b \mod M$ which is the complete key (along with the internal state which is easily calculated given $s_1,g,s_2$). This concludes the cryptanalysis.

\subsubsection{Complexity analysis} \label{carequirements}
Assuming optimization, the total number of calculations is:
\[2^{31}\cdot 2P + 2^{29.4}L + 2^{31}L'\]
In a noise-free scenario, $P=937, L=86, L'=12$, so we get $2^{42}$ calculations. 

In terms of memory, the $\log_2$ table takes $(N-1)\log_2 (N-1)$ bits, but in practice, for better memory access patterns and ease of coding, the attacker keeps a table with $2^{31}$ elements of 32-bit integers, totaling 8GB. The $\log_2 \mod 4$ table takes $2(N-1)$ bits, but again, for better memory access patterns and for ease of coding, the attacker keeps a table of $2^{31}$ elements of one byte (8 bits), totaling 2GB. In all, 10GB RAM storage is needed.

Phases 1-3 are embarrassingly parallelizable (phase 4 has a negligible run time). We implemented them using 48 threads on a 48 physical core machine (on Azure) to get a run time of several minutes, and running the code on a four times stronger machine should result in a real-time attack, i.e., an attack that takes less than one minute to conclude (the seed life span is three minutes).

\section{Impractical Attacks}\label{app:impracticalattacks}

\subsection{A Brute Force Attack on \texorpdfstring{$s_1$}{s1} and \texorpdfstring{$g$}{g}}
The cryptosystem can be attacked without the race-condition using a brute-force enumeration on all $(s_1,g)$ combinations. The attack can proceed with the elimination part of phase 2. An additional enumeration over the three least significant bits of $s_2$ is needed, but this is compensated by the optimization which saves  three  bits of enumeration. So the total enumeration needed to complete phase 2 is $2^{31} \cdot 2^{29.4} = 2^{60.4}$. The total number of calculations required is at least $2^{60.4}L$ which is at least $2^{66.8}$. Thus the attack is slower than the software fault injection attack by a factor of $2^{24.8}$. We see therefore that the brute force attack is impractical.

\subsection{Exploiting Natural Collisions}\label{app:collisions}
\subsubsection{Natural collisions}
While the internal state cannot repeat itself during the lifetime of a seed (the internal state is guaranteed to have a maximal cycle of length $M$, and it is guaranteed not to make more than 1,000,000,000 steps, which is $<M$), two internal states can produce the same PRNG output. This is due to the fact that the internal state is first projected into a space of size $2^{31}$ (via the XOR with $s_2$), and then folded into a space of size $N-1$ via exponentiation (modulo $N$). Specifically, the following pairs of values (after the XOR is applied) are mapped into the same output: $(0,N-1),(1,N),\ldots,(2^{31}-N,2^{31}-1)$. 

Altogether, there are $2^{31}-N+1=20$ such pairs. Thus, every key has up to 20 possible internal state un-ordered pairs that yield output collision. We say ``up to'' because every pair can be mapped to a single internal state pair, or none at all, depending on $s_2$. To illustrate, $s_2=0$ yields no such pairs, because $M<N-1$.

Assuming the key has all 20 pairs available, and given $R$ PRNG outputs from the same seed, the probability to find at least one collision is $\approx 20(\frac{R}{M})^2$.

\subsubsection{Attack}
Suppose the attacker finds a collision $X$ in the PRNG output. The attacker can assume therefore that the values before exponentiation (but after the XOR) are in $\{(0,N-1),(1,N),\ldots,(2^{31}-N,2^{31}-1)\}$. In other words, the attacker can assume that $X \oplus s_1 = g^i \mod N$ for some $0 \leq i \leq 19$. Note, however, that the attacker does not know $i$.

The attacker needs to go over all possible combinations of $g$ values and $i$ values ($20 \cdot \varphi(N-1)=2^{33.7}$ combinations), and each combination also suggests $s_1=X \oplus g^i$. With these candidates, the attacker can proceed to phase 2's elimination step.

In terms of run-time, the attack consumes $2^{33.7}L$ calculations in step 2, thus $2^{40.1}$ calculations. In this respect then, this is still a practical attack.

In terms of memory consumption, this attack requires (using the $\log_2 \mod N$ table) 8GB RAM, which is very practical.

However, the probability of the attack to succeed is very low. Consider an ideal scenario: a 1Gbit/s Ethernet LAN, with MTU=1500B -- this network can deliver no more than 15,000,000 fragmented IPv6 packets in 180 seconds. Setting $R=15,000,000$ yields a probability for at least one collision to be $\approx\frac{1}{750}$. This assumes that the key ($s_2$) does have 20 pairs. As we show below, this applies to $\approx 71\%$ of the keys, while the remaining $\approx 29\%$ of keys have no pairs at all. So the overall success probability of this technique for a given random seed is $\approx\frac{1}{1056}$. On average, therefore, it would take $3\cdot 1056=3168$ minutes (2.2 days) for the attack to succeed, and throughout this entire time the attacker must force the target device to send data at a 1Gbit/s rate. As such, this attack is impractical.

\subsubsection{Pairs per key}
In order for a pair $(i,i+N-1)$ to be ``valid'' in a key $s_2$, the following must hold: $i \oplus s_2 < M$ and $(i+N-1) \oplus s_2 < M$. We enumerated all ($2^{31}$) values of $s_2$ and counted how many pairs are valid under each key. It turns out that for $2^{31}-M \leq s_2 < M$ (1,525,836,544 values), there are 20 pairs per key, and for the remaining 621,647,104 keys -- there are no pairs per key. This means that for $\approx 29\%$ of the keys, the technique does not apply.

\section{Partial Header File Spoofing}\label{app:headerspoof}
One of our case-studies for the attack is NFS, more specifically we demonstrated that we can spoof the $\nth{2}$ fragment of an NFS WRITE call and inject attacker-controlled code into an .h file. The question at hand is how to turn this into a generic utilitarian capability. While .h files in C++ might contain more than just declarations, in C an attacker would have to rely on a specific extension feature supported by the two mainstream C compilers.

The proposed method for C (which also works for C++) is to take advantage of gcc/clang attribute feature. In C/C++, the keyword \texttt{\_\_attribute\_\_} allows specifying special attributes as part of a symbol declaration.  The \texttt{\textbf{constructor}} attribute causes the function to be invoked automatically before execution enters \texttt{main()} (for an executable) or immediately after the binary is loaded (for a library). The attacker can use it to get his/her code executed (when the code that uses the .h file is compiled and executed) without changing any other function or relying on any specific (and unrecommended) coding style of the target code project, particularly placing code in .h files. In order to avoid linker errors due to multiple \texttt{\#include} of the spoofed header file, we also suggest using the  \texttt{weak} attribute. The \texttt{weak} attribute causes the declaration to be emitted as a weak symbol in linkage rather than a global one. A weak symbol allows the symbol to be overridden by a definition with a stronger symbol linkage elsewhere in the program. In other words, if multiple definitions of the same symbol exist, the linker will choose the strongest, if all the symbols are weak, an arbitrary symbol will be used for linking.

The function declaration which has both attributes can be viewed in \autoref{final_spoofed_header}. 
\begin{lstlisting}[language=C, caption=\textbf{test.h}.final spoofed header, label={final_spoofed_header}]
#include<stdio.h>
void __attribute__((weak)) __attribute__((constructor)) 
attacker_controlled_func()
{
    printf("attacker code here, running before main\n");
}
\end{lstlisting}
\section{Cryptanalysis Input Size}\label{app:inputsize}
Below we calculate the size of the data sent from the attacker's frontend device, to the attacker's backend cryptanalysys engine. This size, divided by the bandwidth between the backend and the frontend, determines the communication time overhead which should ideally be negligible, in order not to shorten the time window of the spoofing phase (from the availability of the predicted next fragment IDs, until the next reseed event).

The data required for the cryptanalysis (and the prediction of the next fragment IDs) is as follows (all data pertains to fragment IDs received from the macOS host in the fault triggering phase):

\begin{enumerate}

    \item $P$ sequences of $\mathit{XYZ}$.
    \item The first $L+1$ IDs.
    \item The last ID.
\end{enumerate}

Since each ID consists of four bytes, assuming $L=600$ and an upper bound of $P=3000$, we have a total of $(3000\times3)+(600+1)+1=9602$ IDs which amounts to 38408 bytes (roughly 38.5KB). In our experiments, we send more data by including the duplicate $Y$ (i.e. $\mathit{XYZY}$ instead of $\mathit{XYZ}$) and the sender index of each fragment (four bytes each). Additionally, we send the data as decimal numbers (i.e. in ASCII digits), which further increases the effective data size. In all, the total data size in our experiments is approximately 252KB.

Both the compact data representation (38.5KB) and the data representation we use in our experiments (252KB) can be transmitted in under 0.1 second on a 10 Mbps upload link, which is a fairly modest uplink speed nowadays.

\section{Spoofing TCP Fragments}\label{app:tcpecn}

During the reassembly of fragments, the server's network stack validates the consistency of the  ``Explicit Congestion Notification'' (ECN) IPv6 header flag, ensuring that all fragments have the same value. In UDP, macOS always sends ECN=0, thus the attacker should have ECN=0 in the spoofed fragment.

In TCP, however, it is more complicated, because in general, TCP does support ECN. Hence, the attacker must ensure that the ECN flag in the spoofed $\nth{2}$ fragment matches the ECN flag in the genuine $\nth{1}$ fragment. On macOS (as a TCP client), the value of the IPv6 ECN flag (for packets carrying non-empty TCP payload) in a TCP session is determined by the ``ECN Echo'' (ECE) flag in the TCP header received from the server in the SYN/ACK message (during the TCP handshake). If the server's ECE value is predictable (for example, in the case of a Linux server), the attacker can send a TCP SYN packet to the server before the attack to find out the ECE returned by the server, and thus determine how to set the ECN flag in the spoofed fragment.

However, a macOS server does not have a deterministic ECE value. Fortunately for the attacker, there is a workaround. macOS server only reject fragments when $\text{ECN}\ne 0$ for the $\nth{1}$ fragment and $\text{ECN}=0$ for any other fragment. Taking advantage of this behavior, the attacker can set $\text{ECN}=2$ in the spoofed fragment, and the packet will reassemble correctly.
\section{HTTP Experiment}\label{app:http}
\subsection*{Step 1: Experiment Setup}
The experiment involves four participants.
\begin{enumerate}
    \item Attacker --  The attacker in the experiment is comprised of two logically separate parts, which on our setup are also physically separated. 
    \begin{enumerate}
        \item Frontend -- The attacker device is connected to the switched-LAN, henceforth  we will use the term ``attacker'' or ``attacker device'' to denote the frontend.
        \item  Backend -- The attacker's cryptanalysis machine on Azure, henceforth we will use the term  ``cryptanalytic backend'' to denote it.
    \end{enumerate}
    \item macOS host (HTTP client) -- A Macbook connected to the same switched-LAN as the attacker device. The macOS host runs an echo server listening on UDP port 3333 as a model of the ``fragmentation service''. 
    \item  HTTP server -- A second Macbook (the operating system of the server is immaterial to the attack, in our case we happened to use another Macbook for the server) connected to the same switched-LAN as the attacker device. The HTTP service listens on TCP port 8000.
\end{enumerate}

 \subsection*{Steps 2 - 5}
The same as the NFS attack.

\subsection*{Step 6: Spoofing}
In this step, in contrast to the NFS, the attacker is also required to force the TCP connection between the macOS host and the HTTP server to use fragments. This is done using the technique that is described in \cite[section V.C]{pmtudisnotpanacea}. It is important to note, that in order for the technique to work, the attacker has to win a race, i.e. the attacker has to lower the MTU after the connection is already established and before the HTTP client is sending the actual request. To that end, we insert a short delay after the client is calling  \texttt{connect()}  and before the client calls \texttt{send()}. Another method that will achieve the same goal, is simply to attack a second HTTP request on the same TCP connection.

Next, the attacker uses the known content for the genuine $\nth{2}$ fragment HTTP request to create a spoofed $\nth{2}$ fragment with the modified content  \\\enquote{\texttt{DANGER\_TCP\_IS\_NOT\_SAFE}}. \\
Creating a matching checksum is a bit more involved than what is described in~\autoref{sec:checksum} because the attacker can not use any word (two aligned bytes). For example, non-ASCII bytes, or characters with special meaning (depending on the designated request encoding) may cause the request to be rejected by the application/framework if they are not encoded. To work around this, the attacker can simply use four spare words\footnote{It turns out that all possible sums (modulo $2^{16}-1$) can be achieved with four words of two alphanumeric bytes.} with only alphanumeric characters -- this is equivalent to the one spare word the attacker originally used for the checksum adjustment.
The attacker then sends four copies of the spoofed fragment each with one of the four fragment IDs he/she computed in the previous state. It should also be noted that there is another subtlety when spoofing fragments in TCP which we describe in Appendix ~\ref{app:tcpecn}.

Assuming no other organic fragments are received, there is a 60 seconds window starting at the time the spoofed $\nth{2}$ fragment is sent by the attacker (more accurately -- received by the server) for the genuine request. Next, the macOS HTTP client (running on the host) establishes a TCP connection to the HTTP server. While the HTTP client process is awaiting for the connection to establish, the attacker tricks the macOS host to lower the MTU for the TCP connection. Then, the HTTP client emits an HTTP request as described in~\autoref{httprequest}. The request saves the given token in the server. The token is merely an abstract example, which can be substituted with a more realistic scenario. For example, consider an HTTP request that sets/changes the password of the user where the token is the new password, or an HTTP-based financial transaction where the token is the the amount of money to transfer and the target account. If the macOS sends this genuine request, and the genuine $\nth{1}$ fragment is received at the server during the said 60 seconds time window, the attack will succeed, because the server will immediately match the newly arrived genuine \nth{1} fragment with the spoofed \nth{2} fragment (still in the cache), and the genuine \nth{2} fragment that arrives shortly afterward ``loses the race'' and is discarded.

\begin{lstlisting}[caption={HTTP request}, label={httprequest}]
POST /save HTTP/1.1
Content-Type: application/x-www-form-urlencoded
Content-Length: 1506
Connection: keep-alive

token=RANDOM_1500_ALPHA-NUMERIC_CHARS
\end{lstlisting}

\subsection*{Step 7: Attack success verification}
To verify the success of the attack, we query the HTTP server for the saved token. If the attack succeeds the token should contain our injected string. Again the token here is merely an example, and could be replaced with a more concrete example.

\clearpage
\section{Disassembly}
\subsection{XNU \texttt{ip6\_randomid()} x64}
\begin{lstlisting}[basicstyle=\ttfamily\tiny, caption={XNU \texttt{ip6\_randomid()} x64 disassembly},label={xnu_ipv6_disasm}]
r_L_inline_LCG: 
    mov     esi, cs:randomtab32.ru_a
    mov     edi, cs:randomtab32.ru_b
    mov     ecx, cs:randomtab32.ru_m
    mov     eax, cs:randomtab32.ru_x
; -------------------------------------------------------------------------------
; tmp_ru_x = (randomtab32.ru_b + randomtab32.ru_a * randomtab32.ru_x) 
;            % randomtab32.ru_m;
; -------------------------------------------------------------------------------
    imul    rax, rsi
    add     rax, rdi
    xor     edx, edx
    div     rcx
    test    ebx, ebx
    jz      short r_L_assign_global_PRNG_state ;
; -------------------------------------------------------------------------------
; tmp_ru_x = (randomtab32.ru_b + randomtab32.ru_a * tmp_ru_x) 
;            % randomtab32.ru_m;
; -------------------------------------------------------------------------------
    imul    rdx, rsi
    add     rdx, rdi
    mov     rax, rdx
    xor     edx, edx
    div     rcx
    cmp     ebx, 1
    jz      short r_L_assign_global_PRNG_state ;
; -------------------------------------------------------------------------------
; tmp_ru_x = (randomtab32.ru_b + randomtab32.ru_a * tmp_ru_x) 
;            % randomtab32.ru_m;
; -------------------------------------------------------------------------------
    imul    rdx, rsi
    add     rdx, rdi
    mov     rax, rdx
    xor     edx, edx
    div     rcx
    cmp     ebx, 2
    jz      short r_L_assign_global_PRNG_state ;
; -------------------------------------------------------------------------------
; tmp_ru_x = (randomtab32.ru_b + randomtab32.ru_a * tmp_ru_x)
;            % randomtab32.ru_m;
; -------------------------------------------------------------------------------
    imul    rdx, rsi
    add     rdx, rdi
    mov     rax, rdx
    xor     edx, edx
    div     rcx
r_L_assign_global_PRNG_state: 
; -------------------------------------------------------------------------------
; randomtab32.ru_x = tmp_ru_x
; -------------------------------------------------------------------------------
    mov     cs:randomtab32.ru_x, edx ;
; -------------------------------------------------------------------------------
; randomtab32.ru_counter += n + 1
; -------------------------------------------------------------------------------
    mov     eax, cs:randomtab32.ru_counter
    lea     eax, [rbx+rax+1]
    mov     cs:randomtab32.ru_counter, eax
\end{lstlisting}
\clearpage
\subsection{XNU \texttt{ip6\_randomid()} arm64e}
\begin{lstlisting}[basicstyle=\ttfamily\tiny, caption={XNU \texttt{ip6\_randomid()} arm64e disassembly},label={xnu_ipv6_disasm_arm64e}]
r_L_inline_LCG: 
    mov     esi, cs:randomtab32.ru_a
    mov     edi, cs:randomtab32.ru_b
    mov     ecx, cs:randomtab32.ru_m
    mov     eax, cs:randomtab32.ru_x
; -------------------------------------------------------------------------------
; tmp_ru_x = (randomtab32.ru_b + randomtab32.ru_a * randomtab32.ru_x) 
;            % randomtab32.ru_m;
; -------------------------------------------------------------------------------
    imul    rax, rsi
    add     rax, rdi
    xor     edx, edx
    div     rcx
    test    ebx, ebx
    jz      short r_L_assign_global_PRNG_state ;
; -------------------------------------------------------------------------------
; tmp_ru_x = (randomtab32.ru_b + randomtab32.ru_a * tmp_ru_x) 
;            % randomtab32.ru_m;
; -------------------------------------------------------------------------------
    imul    rdx, rsi
    add     rdx, rdi
    mov     rax, rdx
    xor     edx, edx
    div     rcx
    cmp     ebx, 1
    jz      short r_L_assign_global_PRNG_state ;
; -------------------------------------------------------------------------------
; tmp_ru_x = (randomtab32.ru_b + randomtab32.ru_a * tmp_ru_x) 
;            % randomtab32.ru_m;
; -------------------------------------------------------------------------------
    imul    rdx, rsi
    add     rdx, rdi
    mov     rax, rdx
    xor     edx, edx
    div     rcx
    cmp     ebx, 2
    jz      short r_L_assign_global_PRNG_state ;
; -------------------------------------------------------------------------------
; tmp_ru_x = (randomtab32.ru_b + randomtab32.ru_a * tmp_ru_x)
;            % randomtab32.ru_m;
; -------------------------------------------------------------------------------
    imul    rdx, rsi
    add     rdx, rdi
    mov     rax, rdx
    xor     edx, edx
    div     rcx
r_L_assign_global_PRNG_state: 
; -------------------------------------------------------------------------------
; randomtab32.ru_x = tmp_ru_x
; -------------------------------------------------------------------------------
    mov     cs:randomtab32.ru_x, edx ;
; -------------------------------------------------------------------------------
; randomtab32.ru_counter += n + 1
; -------------------------------------------------------------------------------
    mov     eax, cs:randomtab32.ru_counter
    lea     eax, [rbx+rax+1]
    mov     cs:randomtab32.ru_counter, eax
\end{lstlisting}

\end{document}